\documentclass[
twocolumn
]{aastex631}

\usepackage{amsmath, subfigure}
\usepackage{threeparttable}
\usepackage{orcidlink}
\usepackage{url}

\begin{document}
\begin{sloppypar}

\title{The Disk Wind Contribution to the Gamma-Ray emission from the nearby Seyfert Galaxy GRS~1734-292}

\author[0009-0004-5978-1785]{Nobuyuki Sakai}
\affiliation{Department of Earth and Space Science, Graduate School of Science, Osaka University, 1-1 Machikaneyama, Toyonaka, Osaka 560-0043, Japan}

\author{Tomoya Yamada}
\affiliation{Department of Earth and Space Science, Graduate School of Science, Osaka University, 1-1 Machikaneyama, Toyonaka, Osaka 560-0043, Japan}

\author[0000-0002-7272-1136]{Yoshiyuki Inoue}
\affiliation{Department of Earth and Space Science, Graduate School of Science, Osaka University, 1-1 Machikaneyama, Toyonaka, Osaka 560-0043, Japan}
\affiliation{Interdisciplinary Theoretical \& Mathematical Science Program (iTHEMS), RIKEN, 2-1 Hirosawa, 351-0198, Japan}
\affiliation{Kavli Institute for the Physics and Mathematics of the Universe (WPI), UTIAS, The University of Tokyo, 5-1-5 Kashiwanoha, Kashiwa, Chiba 277-8583, Japan}

\author[0000-0003-1052-6439]{Ellis R. Owen}
\affiliation{Department of Earth and Space Science, Graduate School of Science, Osaka University, 1-1 Machikaneyama, Toyonaka, Osaka 560-0043, Japan}

\author[0000-0003-2475-7983]{Tomonari Michiyama}
\affiliation{Faculty of Information Science, Shunan University, 843-4-2 Gakuendai, Shunan, Yamaguchi 745-8566, Japan}

\author[0000-0002-6797-2539]{Ryota Tomaru}
\affiliation{Department of Earth and Space Science, Graduate School of Science, Osaka University, 1-1 Machikaneyama, Toyonaka, Osaka 560-0043, Japan}

\author[0000-0002-0921-8837]{Yasushi Fukazawa}
\affiliation{Department of Physical Science, Hiroshima University, 1-3-1 Kagamiyama, Higashi-Hiroshima, Hiroshima 739-8526, Japan}
\affiliation{Hiroshima Astrophysical Science Center, Hiroshima University, 1-3-1 Kagamiyama, Higashi-Hiroshima, Hiroshima 739-8526, Japan}
\affiliation{Core Research for Energetic Universe (Core-U), Hiroshima University, 1-3-1 Kagamiyama, Higashi-Hiroshima, Hiroshima 739-8526, Japan}

\correspondingauthor{Nobuyuki Sakai}
\email{u938638f@ecs.osaka-u.ac.jp}

\begin{abstract}
Radio-quiet Seyfert galaxies have been detected in GeV gamma-rays by the \textit{Fermi} Large Area Telescope (LAT), but the origin of much of this emission is unclear. We consider the nearby example GRS 1734-292, which exhibits weak starburst and jet activities that are insufficient to explain the observed gamma-ray flux. With the first detailed multi-wavelength study of this source, we demonstrate that an active galactic nucleus (AGN) disk wind can account for its gamma-ray emission. Using a lepto-hadronic emission model based on a shocked ambient medium and a shocked wind region created by an AGN accretion disk wind, we identify two viable scenarios that are consistent with the \textit{Fermi}-LAT data and multi-wavelength observations: a hadronic $pp$-dominated scenario and a leptonic external Compton-dominated scenario. Both of these show that future observations with the Cherenkov Telescope Array (CTA) and the Southern Wide-field Gamma-ray Observatory (SWGO) could detect TeV emission from a disk wind in GRS~1734-292. Such a detection would substantially improve our understanding of cosmic ray acceleration efficiency in AGN disk wind systems, and would establish radio-quiet Seyfert galaxies as cosmic ray accelerators capable of reaching ultra-high energies. 

\end{abstract}


\section{Introduction} \label{sec:intro}
The \textit{Fermi} Large Area Telescope (LAT) has detected 6,658 objects over an energy range of 50~MeV to 1~TeV \citep{Abdollahi+22}. More than half are extragalactic and, of these, most have been classified as blazars. Recently, \textit{Fermi}-LAT  
also reported the detection of gamma-ray emission from Seyfert galaxies. These are radio-quiet active galactic nuclei (AGN) that lack strong relativistic jets \citep{Abdollahi+20, Abdollahi+22}. NGC 1068 is one example of this source type~\citep{Ajello:2023hkh} which has also been identified as a TeV neutrino source with a significance of 4.2\text{-}$\sigma$ \citep{IceCube+22}. While starburst activities \citep[e.g.,][]{Lenain+2010, Eichmann:2022lxh, Ajello:2023hkh}, weak jets \citep[e.g.,][]{Inoue:2023bmy, Fang+2024, Salvatore:2023zmf, Yasuda:2024fvc}, coronae \citep[e.g.,][]{Inoue:2019fil, Murase:2019vdl}, and disk winds \citep[e.g.,][]{Lamastra+16, Wand&Loeb16a, Wang&Loeb16b, Liu+18, Inoue+22, Peretti+23, Huang:2024yua} have been proposed as potential sources of the high-energy emission from Seyfert galaxies, its exact origin remains unsettled. 

GRS 1734-292 is another example of a Seyfert galaxy detected by \textit{Fermi}-LAT. In this system, the contribution to the GeV gamma-ray emission from starburst activities and a jet are empirically constrained to be small (see Section \ref{sec: GRS}), and coronae are generally not able to emit GeV gamma-rays due to internal attenuation \citep{Inoue:2019fil}. Additionally, \citet{Ajello+21} reported a detection of GeV gamma-rays at $5.1\text{-}\sigma$ confidence level from disk winds associated Seyfert galaxies using a stacking analysis of 11.1~years of \textit{Fermi}-LAT survey data. These observations suggest a disk wind is the likely origin of GeV gamma-ray emission from GRS 1734-292 and indicate that this system is a natural laboratory to test the disk wind scenario without significant contamination from other emission components. 

In this work, we calculate the lepto-hadronic emission from high energy cosmic rays (CRs) accelerated at shock fronts created by a disk wind interacting with the interstellar medium (ISM) of GRS~1734-292. We adopt the disk wind model described by \citet{Yamada+24} and compare our results with multi-wavelength data of GRS~1734-292. This work is the first to compare a theoretical model with multi-wavelength observations of this source. Throughout this paper, we adopt standard cosmological parameter values of $(h, \Omega_M, \Omega_\Lambda) = (0.7, 0.3, 0.7)$.

\section{Observed Properties of GRS~1734-292}
\label{sec: GRS}

GRS~1734-292 is a radio-quiet Seyfert galaxy at a redshift of 0.0214 \citep[93~Mpc;][]{Marti+98}. It has a bolometric disk luminosity of $L_\mathrm{AGN}=1.45\times 10^{45}~\mathrm{erg~ s^{-1}}$ \citep{Tortosa+17}, and was detected by \textit{Fermi}-LAT with a significance of $4.93\text{-}\sigma$, a 
gamma-ray spectral index of $2.38\pm0.28$, and a flux of $(1.16\pm0.32)\times10^{-11}\ \mathrm{erg\ cm^{-2}\ s^{-1}}$ reported between 0.1--100~GeV~\citep{Ballet+2023}. This corresponds to a gamma-ray luminosity of $\sim10^{43}~\mathrm{erg~s^{-1}}$.

The starburst and jet activities in GRS~1734-292 are empirically insufficient to account for its observed gamma-ray flux. Its total infrared (IR) luminosity is less than $3\times10^{10}\ L_\odot$ \citep{Shimizu+16}, corresponding to a gamma-ray luminosity of $<3\times10^{40}~\mathrm{erg~s^{-1}}$ if following empirical relations between IR and gamma-ray luminosities of star-forming galaxies \citep{Ajello+20}. The 1.4~GHz radio luminosity is $7\times10^{22}~\mathrm{W~Hz^{-1}}$ \citep{Marti+98}, which would correspond to a gamma-ray luminosity of $\sim10^{41}~\mathrm{erg~s^{-1}}$ if following correlations between radio and gamma-ray luminosities \citep{Inoue:2011bm, Fukazawa+22}.

In Galactic coordinates, GRS~1734-292 is located at $l=358.89^\circ$ and $b=1.41^\circ$. This is close to the Galactic center so UV-optical fluxes from this source are expected to be strongly attenuated by Galactic extinction. We model the UV-optical spectrum of GRS 1734-292 using a multi-color black body with an inner accretion disk temperature of $\sim12$~eV and a luminosity of $\approx2\times10^{43}~\mathrm{erg~s^{-1}}$ in the $4900\text{--}9000\ \mathrm{\AA}$ band \citep{Marti+98}. This reproduces the bolometric luminosity $L_\mathrm{AGN}=1.45\times10^{45}\ \mathrm{erg\ s^{-1}}$. In the IR band, the \citet{https://doi.org/10.26132/ned1} archival data of GRS~1734-292 shows two spectral bumps at $\approx0.3$~eV ($10^3$~K) and $\approx0.03$~eV ($10^2$~K). We consider that these IR bumps originate from AGN activity, where typical temperatures of the narrow-line region and torus are $\sim10^2\ \mathrm{K}$ and $\sim10^3\ \mathrm{K}$, respectively \citep[e.g.,][]{2015ARA&A..53..365N}. 
We use the X-ray
observations of \citet{Tortosa+17} to complete our multi-wavelength view of GRS~1734-292. 
The spectral energy distribution (SED) of the modeled AGN disk is shown together with the multi-wavelength data in Figure~\ref{fig:sed}.

Recently, \citet{Michiyama+24} reported mJy-level millimeter radio flux. However, given the observed day-scale variability, the origin of the millimeter radio emission in GRS~1734-292 is likely to be the compact corona rather than the extended disk wind. We therefore treat their data as upper limits here.

\section{Lepto-hadronic Emission from an AGN Disk Wind}
\label{sec:Method} 

We follow \citet{Yamada+24}  to model the disk wind interaction with the ISM (see also \citealt{Andre+12} and \citealt{Nims+15}). This invokes a fast disk wind, also known as an ultrafast outflow \citep[UFO, e.g.,][]{Tombesi+15, Mizumoto+19}, launched from the inner accretion disk, which interacts with the ISM. The velocity of a UFO is typically  $v_\mathrm{UFO}\sim0.03\text{--}0.3c$, where $c$ is the speed of light. We adopt the maximum value in this range ($0.3c$) in our calculations. UFO features are yet to be detected in the X-ray spectrum of GRS~1734-292. 
Outflow structures are angle-dependent, and their detectability depends on the line-of-sight to the observer so the non-detection of UFO features in GRS~1734-292 does not 
imply they are not present~\citep[e.g.,][]{Nomura+16}. A fraction \( f_\mathrm{UFO} \) of the radiation momentum is converted into the kinetic momentum of the UFO as $\dot{M}_\mathrm{UFO} v_\mathrm{UFO} = f_\mathrm{UFO} {L_\mathrm{AGN}}/{c}$, where $\dot{M}_\mathrm{UFO}$ is the mass outflow rate of the UFO. We model the density profile of the ISM as $n_\mathrm{ISM}(r) = n_0 ({r}/{r_0})^{-\beta}$, where \( n_0 \) is the number density of ISM protons at radius \( r_0 \). Here, we set \( r_0 = 100\ \mathrm{pc} \) and \( \beta = 1\) \citep{Nims+15}. 
  
The interaction between the disk wind and the ISM creates a forward shock (FS) and a reverse shock (RS). These shocks generate a shocked ambient medium (SAM) and a shocked wind (SW) region, which are separated by a contact discontinuity \citep{Koo&McKee1992ApJ...388...93K, Koo&McKee1992b, Andre+12, Nims+15, Yamada+24}. The radii of the forward shock and reverse shock at an age of $t_\mathrm{wind}$ during the fully adiabatic stage are estimated as \citep{Koo&McKee1992ApJ...388...93K, Koo&McKee1992b}:
\begin{align}
    R_\mathrm{FS}(t_\mathrm{wind}) &=A_\mathrm{FS}\left(\frac{L_\mathrm{wind}}{\mu_\mathrm{H}m_pn_0r_0^\beta}\right)^{\frac{1}{5-\beta}}t_\mathrm{wind}^{\frac{3}{5-\beta}}\label{eq: R_FS}\\
    &= \left(\frac{2\pi}{3-\beta}\right)^{\frac{1}{5-\beta}}A_\mathrm{FS}v_\mathrm{UFO} \tau_\mathrm{free} \left( \frac{t_\mathrm{wind}}{\tau_\mathrm{free}} \right)^{\frac{3}{5 - \beta}}, \\
    R_\mathrm{RS}(t_\mathrm{wind}) &=A_\mathrm{RS}\left(\frac{L_\mathrm{wind}}{\mu_\mathrm{H}m_pn_0r_0^\beta}\right)^{\frac{3}{2(5-\beta)}}v_\mathrm{UFO}^{-\frac12}t_\mathrm{wind}^{\frac{4+\beta}{2(5-\beta)}}\label{eq: R_RS}\\ 
    &= \left(\frac{2\pi}{3-\beta}\right)^{\frac{3}{2(5-\beta)}}A_\mathrm{RS}v_\mathrm{UFO} \tau_\mathrm{free} \left( \frac{t_\mathrm{wind}}{\tau_\mathrm{free}} \right)^{\frac{4 + \beta}{2(5 - \beta)}}
\end{align}
for $t_\mathrm{wind}>\tau_\mathrm{free}$, where
\begin{align}
    \tau_\mathrm{free} = \frac{r_0}{v_\mathrm{UFO}}\left[\frac{(3-\beta)\dot M_\mathrm{UFO}}{4\pi r_0^2v_\mathrm{UFO}\mu_\mathrm{H}m_pn_0}\right]^{\frac{1}{2-\beta}}
\end{align}
is the free-expansion timescale, $L_\mathrm{wind}\equiv\dot{M}_\mathrm{UFO}v_\mathrm{UFO}^2/2$ is the kinetic power of the disk wind, $\mu_\mathrm{H}=1.4$ is the mean molecular weight per particle, and $m_p$ is the mass of a proton. $A_\mathrm{FS}$ and $A_\mathrm{RS}$ are dimensionless constants in the order of unity and based on \citet{Koo&McKee1992b}. They are given as:
\begin{align}
    A_\mathrm{FS} &\approx \left[\frac{2(3-\beta)(5-\beta)^3}{9\pi(11-\beta)(7-2\beta)\lambda_\mathrm{CD}^3}\right]^{\frac{1}{5-\beta}}
\end{align}
and
\begin{align}
    A_\mathrm{RS} &\approx 2 \left(\frac{16}{15}\right)^{\frac34} \left[\frac{11-\beta}{5(5-\beta)}\right]^{\frac12} (\lambda_\mathrm{CD} A_\mathrm{FS})^{\frac32},
\end{align}
where $\lambda_\mathrm{CD} \equiv R_\mathrm{CD} / R_\mathrm{FS} \approx (123 - 8\beta) / (143 - 8\beta)$ is the ratio of the radius of the contact discontinuity $R_\mathrm{CD}$ to the forward shock radius $R_\mathrm{FS}$.

At the forward and reverse shock fronts, charged particles are accelerated through diffusive shock acceleration \citep[DSA;][]{Drury+83, Bell78a, Bell78b, Blandford&Ostriker78}. The acceleration timescale of a CR with energy of $E_\mathrm{CR}$ is $\tau_\mathrm{CR,acc}=8\eta_{\mathrm g}cE_\mathrm{CR}/(3eBv_\mathrm{up}^2)$, where $\eta_\mathrm{g}$ is the gyro-factor of the system, $e$ is the elementary charge, $B$ is the magnetic field in the upstream region, and $v_\mathrm{up}$ is the upstream velocity in the shock rest frame \citep{Drury+83,Yamada+24}. To model the CR spectrum, ${dN_\mathrm{CR}}/{dE_\mathrm{CR}}$, 
we consider that the system is approximately steady over an observational time ($\sim10\ \mathrm{yr}$). This is justified, as the shock and cooling properties change very little over a 10 year timescale, which  
is much shorter than 
the evolutionary timescale of the system we consider
($\sim10^{3\text{--}5}\ \mathrm{yr}$, c.f. a typical AGN phase lifetime, e.g., \citealt{2015MNRAS.451.2517S}) and 
generally shorter than 
the CR cooling timescales over most the energy range we consider (c.f. Figure \ref{fig:timescales_pp} and \ref{fig:timescales_EC}).
This allows us to write the transport equation as: 
    \begin{align}
        \frac{d}{dE_\mathrm{CR}} \left( \dot{E}_\mathrm{cool} \frac{dN_\mathrm{CR}}{dE_\mathrm{CR}} \right) + \frac{1}{\tau_\mathrm{adv}(t_\mathrm{wind})} \frac{dN_\mathrm{CR}}{dE_\mathrm{CR}} = Q_\mathrm{CR}(E_\mathrm{CR}),
        \label{eq:kinetic_eq}
    \end{align}
where \( \dot{E}_\mathrm{cool} \) is the cooling rate of CRs, $\tau_\mathrm{adv}(t_\mathrm{wind})$ is the advection timescale of the CRs at $t_\mathrm{wind}$, and \( Q_\mathrm{CR}(E_\mathrm{CR}) \) is the injection rate of CRs with an energy of $E_\mathrm{CR}$.
Here, electron cooling is dominated by synchrotron and inverse Compton losses \citep{Blumenthal&Gould70}. By contrast, hadronuclear ($pp$) interactions, photopion production, and Bethe-Heitler processes dominate the losses of protons \citep[]{Kelner+06, Kelner+08} where target photon fields are comprised of internal synchrotron emitted photons, external AGN disk photons, and the cosmic microwave background. The magnetic field of the SAM region is parameterized as $B_\mathrm{SAM}$, while that of the SW is defined by the magnetic energy fraction in the UFO wind $\epsilon_{B, \mathrm{SW}}$. Additionally, as SWs expand adiabatically, we account for the adiabatic losses experienced by the entrained particles \citep{Koo&McKee1992ApJ...388...93K, Koo&McKee1992b, Andre+12}. 
We set $\tau_\mathrm{adv}(t_\mathrm{wind})=R_\mathrm{FS}/V_\mathrm{FS}$ for the SAM and $\tau_\mathrm{adv}(t_\mathrm{wind})=R_\mathrm{RS}/(v_\mathrm{UFO}/4+V_\mathrm{RS})$ for the SW, where $V_\mathrm{FS/RS}\equiv dR_\mathrm{FS/RS}/dt_\mathrm{wind}$ is the velocity of the forward/reverse shock. We adopt an exponential cutoff power-law to model the injected particle spectrum 
$Q_\mathrm{CR}(E_\mathrm{CR}) \propto E^{-p_\mathrm{CR}} \exp\left( -{E_\mathrm{CR}}/{E_\mathrm{CR,max}} \right)$, 
which is appropriate for DSA \citep[see e.g.,][]{Drury+83}. 
Here, \( p_\mathrm{CR} \) is set to be the same for protons and electrons.

We solve equation~\ref{eq:kinetic_eq} following \citet{Ginzburg+66} to obtain a CR particle spectrum. This is normalized to a fraction of 
the thermal energy injected into the SAM or SW region \( \xi_i L_\mathrm{th, SAM/SW} \) \citep{Wand&Loeb16a, Liu+18}, representing the energy transfer from the shocked thermal gas to the CRs (of species $i = \{ {\rm p}, {\rm e} \}$) as they are accelerated. The thermal energy injected into the SAM is given by \citep{1977ApJ...218..377W, Koo&McKee1992b}
\begin{align}\label{eq: L_th_SAM}
    L_\mathrm{th,SAM} \approx \frac{9(5-\beta)(143-8\beta)^3}{4(11-\beta)(7-2\beta)(123-8\beta)^3}L_\mathrm{wind}
\end{align}
and that into the SW is
\begin{align}\label{eq: L_th_SW}
    L_\mathrm{th,SW} \approx \frac{5-\beta}{11-\beta}L_\mathrm{wind}.
\end{align}
The maximum CR energy $E_\mathrm{CR, max}$ is determined by the balance between the acceleration timescale against the advection or cooling timescale: $\tau_\mathrm{CR,acc}^{-1}=\tau_\mathrm{adv}^{-1}+\tau_\mathrm{cool}^{-1}$, where $\tau_\mathrm{cool}=E_\mathrm{CR}/\dot E_\mathrm{cool}$.
    
High-energy CRs produce photons, neutrinos, and other secondary particles. In our model, we consider the emission of photons and neutrinos associated with synchrotron processes including synchrotron self-absorption \citep{Rybicki+86}, inverse Compton scattering \citep{Khangulyan+14}, $pp$ interactions \citep{Kelner+06}, photopion production \citep{Kelner+08}, and Bethe-Heitler processes \citep[]{Blumenthal70, Kelner+08}. Of these, $pp$ interactions, photopion production, and Bethe-Heitler processes also produce secondary electrons and positrons. These secondaries can also undergo synchrotron and inverse Compton processes to emit photons, and we self-consistently account for their contribution in our model. We consider both synchrotron self-Compton and external Compton (EC) when calculating the inverse Compton emission, where seed photons are supplied by the internal synchrotron and the central AGN activity, respectively. We also take into account gamma-ray attenuation by AGN photons \citep{Inoue:2019fil} and the extragalactic background light \citep{Inoue+13a}.

In this study, we consider two configurations of our AGN disk wind model. The first is a \textit{pp}-dominated scenario, where neutral pion decays formed in $pp$ interactions dominate the GeV gamma-ray emission. The second is an EC-dominated scenario, where external Compton emission dominates at GeV energies.

\begin{table}[t]
\centering
\caption{Parameters of the $pp$-dominated scenario and EC-dominated scenario}
\begin{tabular}{lccc}
    \hline \hline 
    Parameter & $pp$-dominated & EC-dominated \\
    \hline
    $L_{\mathrm{AGN}}\ (\mathrm{erg\ s^{-1}})$ $^{\dagger}$ & $1.45\times10^{45}$ & $1.45\times10^{45}$ \\
    $v_{\mathrm{UFO}}$ & 0.3$c$ & $0.3c$   \\
    $f_{\mathrm{UFO}}$ & 10 & 40 \\
    $t_\mathrm{wind}\ (\mathrm{yr})$ & $10^5$ & $3\times10^3$  \\
    $n_0\ (\mathrm{cm^{-3}})$ & 200 & 10 \\
    $B_\mathrm{SAM}\ (\mathrm{\mu G})$ & $10$ & $10$  \\
    $\epsilon_{B, \mathrm{SW}}$ & $3\times10^{-4}$ & $4\times10^{-5}$ \\
    $p_\mathrm{CR}$ & $2.2$ & $2.2$  \\
    $\xi_p$ & 0.1 & 0.1 \\
    $\xi_e$ & 0.01 & 0.01 \\
    $\eta_\mathrm{g}$ & 1 & 1  \\
    $\tau_\mathrm{free}\ (\mathrm{yr})$ & $2.3\times10^{-2}$ & 1.9 \\
    \hline
\end{tabular}
\\
\raggedright
Note: $^{\dagger}$ from \citet{Tortosa+17}.
\label{tab:fiducial parameter}
\end{table}

\section{Results}
\label{sec:Result}

In both the \textit{pp}-dominated and EC-dominated scenarios, we set the gyro-factor $\eta_\mathrm{g}=1$, which corresponds to the Bohm limit. We also adopt values of 
$\xi_p=0.1,\ \xi_e=0.01$ for the energy fractions of CR protons and electrons, respectively. These choices follow previous studies of AGN disk winds \citep[e.g.,][]{Lamastra+16} which were informed by observations of supernova remnants \citep[e.g.,][]{Ackermann+13}.
The value of $B_\mathrm{SAM}$ is common to both model scenarios, but $\epsilon_{B, \mathrm{SW}}$ is allowed to differ. Their choices are set such that our model reproduces the centimeter radio data without violating constraints from Atacama Large Millimeter/submillimeter Array (ALMA) observations \citep{Michiyama+24}.
Other model parameters, $f_\mathrm{UFO},\ t_\mathrm{wind},\ \mathrm{and}\ n_0$, vary between the two scenarios. 

In the \textit{pp}-dominated scenario, we set $f_\mathrm{UFO}=10$, \(n_0 = 200\ \mathrm{cm^{-3}}\), and $t_\mathrm{wind}=10^{5}\ \mathrm{yr}$, leading to $R_\mathrm{FS}=198~\mathrm{pc}$ and $R_\mathrm{RS}=34\ \mathrm{pc}$. Our choice of density is comparable with levels seen in the Central Molecular Zone of the Milky Way \citep[e.g.,][]{Molinari2011ApJ...735L..33M, Tsuboi2015PASJ...67...90T}. 

In the EC-dominated scenario, we set an age of \(t_\mathrm{wind} = 3\times10^3\ \mathrm{yr}\), leading to smaller radii of $R_\mathrm{FS}=43\ \mathrm{pc}$ and $R_\mathrm{RS}=20\ \mathrm{pc}$, and higher densities of seed photons. We also set $f_\mathrm{UFO}=40$, which remains within the observed range \citep{Mizumoto+19}. For the gas density, we set $n_0=10\ \mathrm{cm}^{-3}$, following \citet{Nims+15} and \citet{Yamada+24}. 
Our parameter choices for each model scenario are summarized in Table~\ref{tab:fiducial parameter}. 

\begin{figure*}[p]
    \centering
    \includegraphics[width=.8\linewidth]{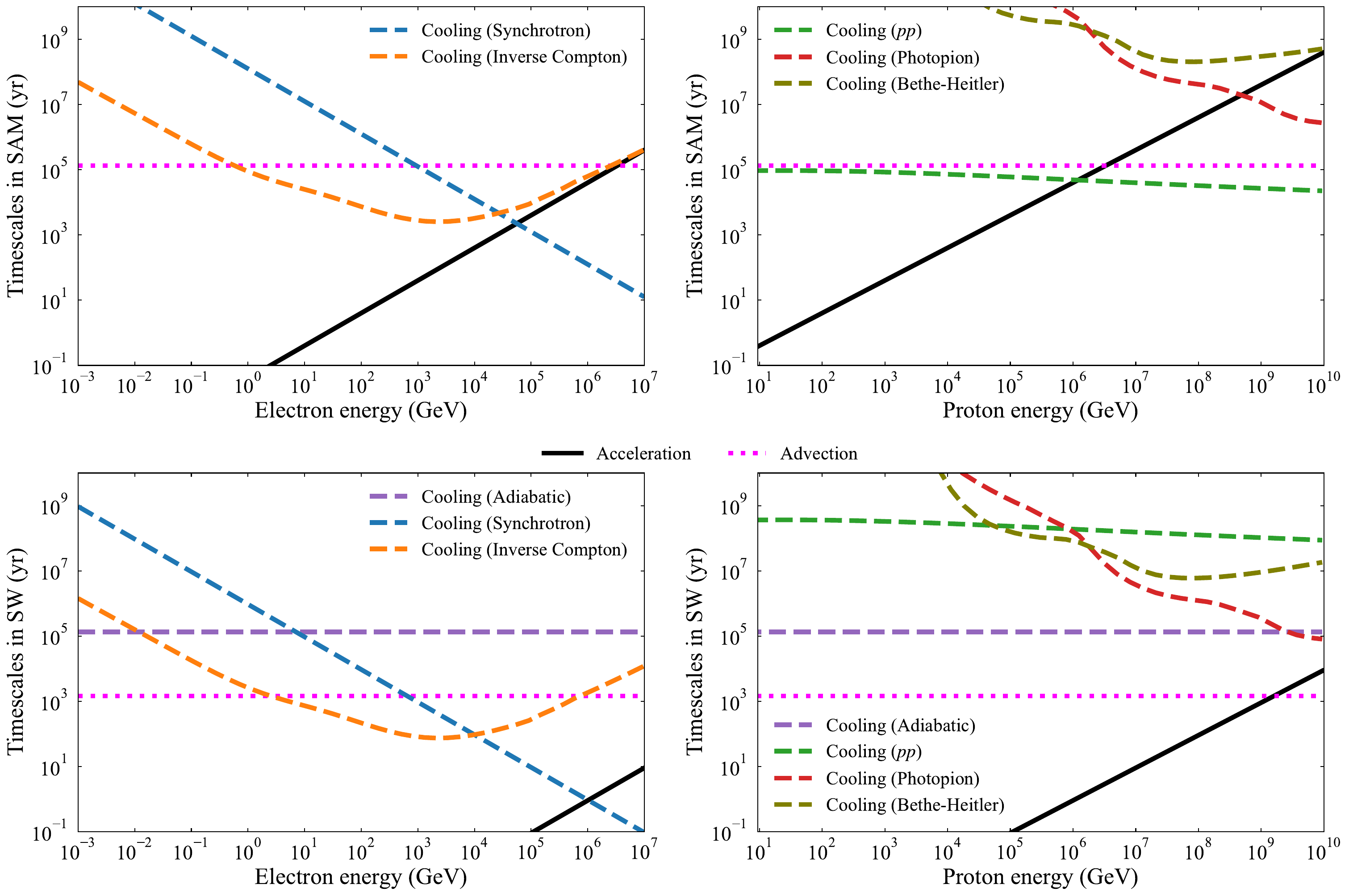}
    \caption{The acceleration, advection, and cooling timescales of CRs in the SAM and SW regions for the \textit{pp}-dominated scenario. The black solid and magenta dotted lines show the acceleration and advection timescales of CRs. The purple, blue, orange, green, red, and olive dashed lines show cooling timescales of adiabatic, synchrotron, inverse Compton scattering, \textit{pp} interactions, photopion production, and Bethe-Heitler processes, respectively. Panels on the upper left, upper right, lower left, and lower right show the timescales of CR electrons in the SAM, CR protons in the SAM, CR electrons in the SW, and CR protons in the SW. The parameters used here are shown in Table \ref{tab:fiducial parameter}.}
    \label{fig:timescales_pp}
\end{figure*}

\begin{figure*}[h]
    \centering
    \includegraphics[width=.8\linewidth]{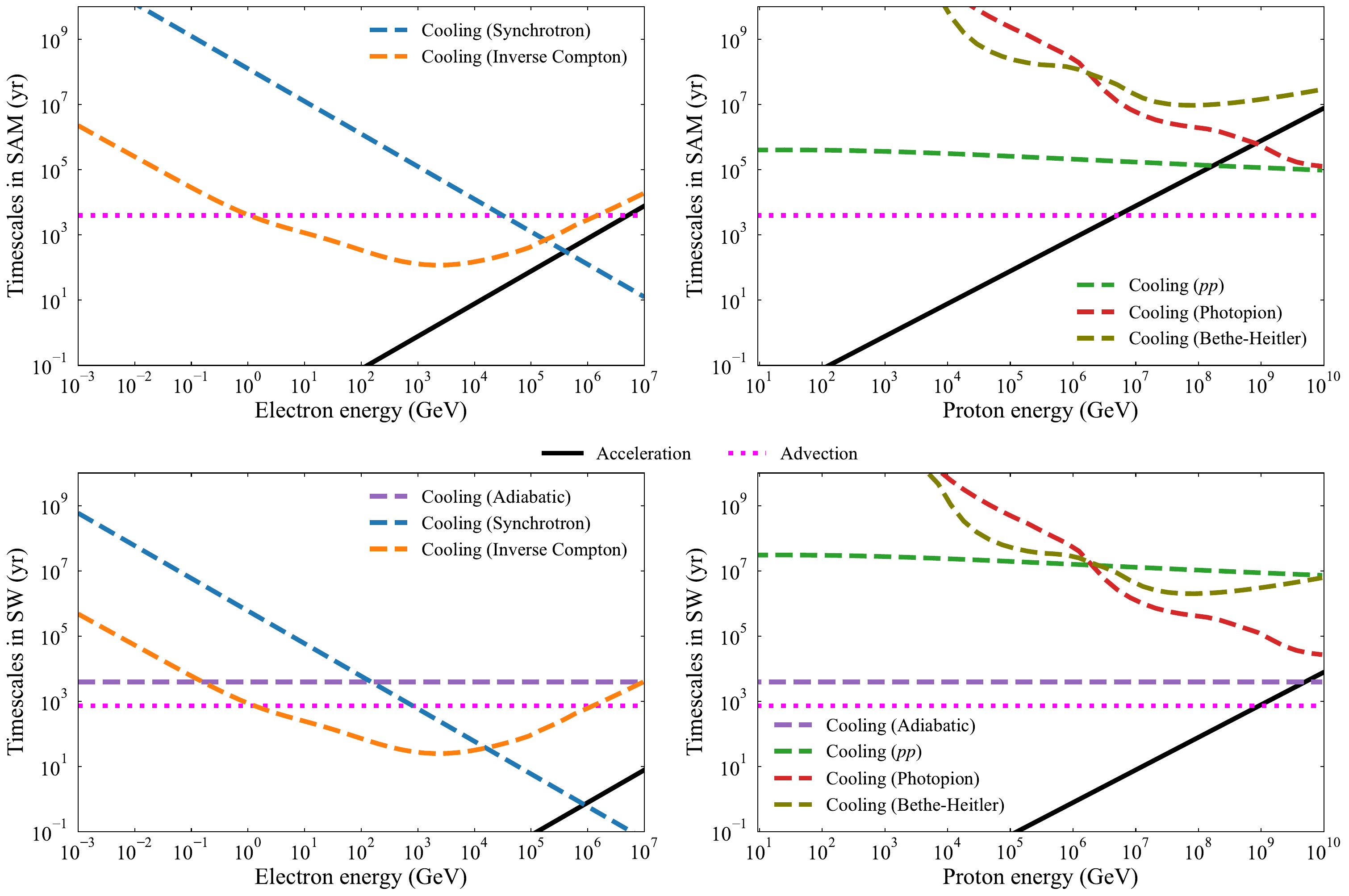}
    \caption{Same as in Figure \ref{fig:timescales_pp}, but for the EC-dominated scenario.}
    \label{fig:timescales_EC}
\end{figure*}

\begin{figure*}[t]
\centering
    \begin{minipage}{0.48\linewidth}
        \centering
        \includegraphics[height=0.235\textheight]{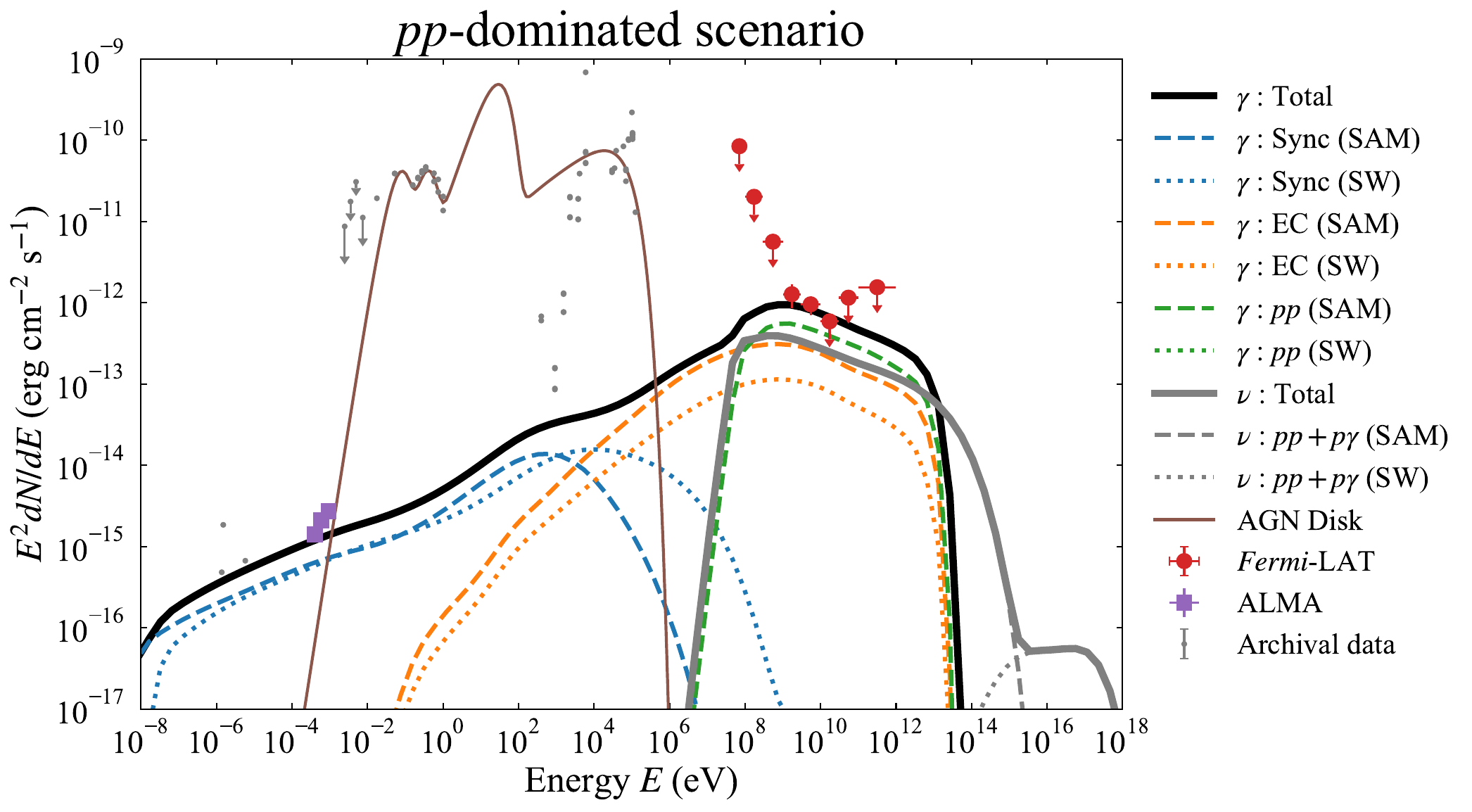}
    \end{minipage}
    \hspace{0.02\linewidth}
    \begin{minipage}{0.48\linewidth}
        \centering
        \includegraphics[height=0.235\textheight]{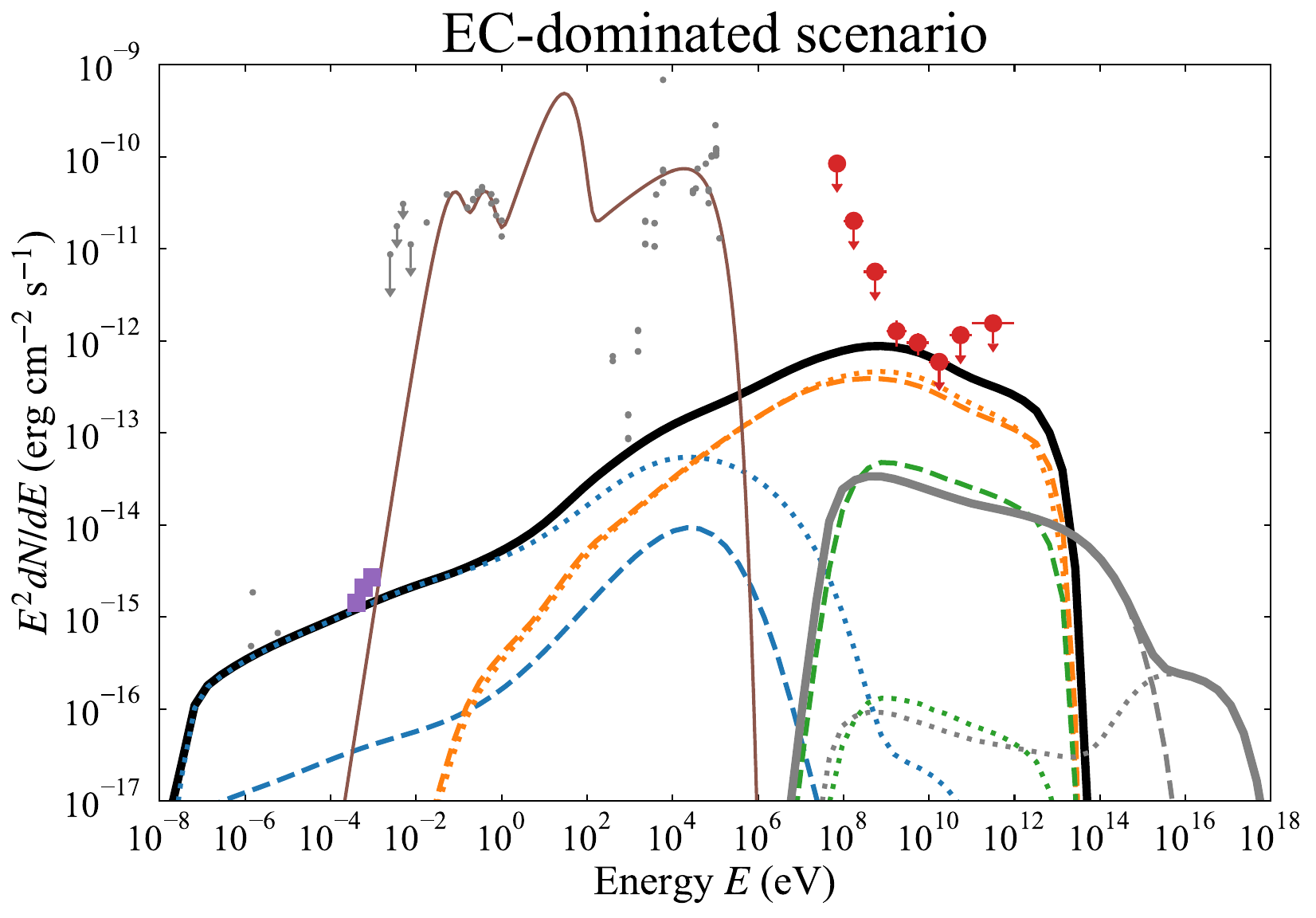}
    \end{minipage}
    \caption{{\it Left}: SED of GRS 1734-292 for the $pp$-dominated scenario. The black and gray solid lines show the total photon and neutrino model spectra. The blue, orange, and green lines show photon spectra from synchrotron, EC, and \textit{pp} interactions. The gray lines show neutrino spectra. The dashed and dotted lines show contributions from the SAM and SW regions, respectively. Red circles, purple squares, and gray dots show data from \textit{Fermi}-LAT, ALMA, and archival data collated from the NED. This includes radio observations conducted with the Very Large Array \citep{Condon+1998, Nord+2004} and the UTRAO  \citep{Douglas+1996}, IR data from the 2MASS all-sky survey  \citep{Skrutskie+2006}, \textit{Herschel} and \textit{Spitzer}/IRAC  \citep{Melendez+2014}, and SPIRE \citep{Shimizu+16}, and X-ray data obtained with \textit{Chandra} \citep{Wang+2016}, \textit{INTEGRAL} \citep{Sazonov+2004, Bird+2007, Sazonov+2007, Beckmann+2009, Malizia+2009, Krivonos+2015}, \textit{Swift}\text{-}BAT \citep{Tueller+2008, Cusumano+2010, Tueller+2010, Ricci+2017ApJS..233...17R, Oh+2018ApJS..235....4O}, and \textit{XMM--Newton} \citep{Boissay+2016A&A...588A..70B}. The brown line shows the AGN disk SED. There are clear differences between the modeled SED and the archival data at $\sim$ 1--10 keV. This is because soft X-ray photons are absorbed by interstellar gas during their propagation, which our model does not explicitly account for.
    {\it Right}: Same as in the left panel but for the EC-dominated scenario.}
\label{fig:sed}
\end{figure*}

Figure \ref{fig:timescales_pp} shows the acceleration, advection, and cooling timescales of CRs for the \textit{pp}-dominated scenario. In both the top and bottom left panels, corresponding to CR electrons in the SAM and SW, it can be seen that electrons in the Thomson regime cool primarily through inverse Compton, but synchrotron cooling becomes more important at higher energies due to the Klein-Nishina effect. The maximum energy of the CR electrons is therefore determined by the balance between the acceleration and the synchrotron cooling timescales. For CR protons in the dense SAM environment, \textit{pp} interactions dominate cooling across all energies (cf. the top right panel). Conversely, in the SW region, advection losses are more efficient than the total cooling for CR protons. This implies a maximum CR proton energy of $\sim10^{18}\ \mathrm{eV}$ can be achieved, given the balance between the acceleration and the advection timescales shown in the bottom-right panel (see also Section \ref{sec:UHECR}).

Figure \ref{fig:timescales_EC} illustrates the same timescales for the EC-dominated scenario. A key distinction in this scenario is that even in the SAM, advection is more efficient for CR protons than the total cooling effect (the top right panel). This follows from the lower gas densities compared to the \textit{pp}-dominated scenario. In the SW, CR protons are also accelerated to energies up to $\sim10^{18}\ \mathrm{eV}$ (the bottom right panel).

Figure~\ref{fig:sed} shows the multi-messenger spectra including photons and neutrinos for the two model scenarios, together with gamma-ray observations of GRS 1734-292 obtained by 
\textit{Fermi}-LAT \citep{Ballet+2023}\footnote{This shows \textit{Fermi}-LAT 4FGL-DR4 data, available online \\ (\url{https://fermi.gsfc.nasa.gov/ssc/data/access/lat/14yr_catalog/}).}, cm-wavelength data from ALMA \citep{Michiyama+24}, and archival radio, IR, and X-ray data collated from NED. The total contributions to the photon and neutrino emission from primary and secondary CRs are shown. The contributions from the SAM and SW regions are distinguished by line styles.

In both the \textit{pp}-dominated and EC-dominated scenarios, we find that the photopion contribution to the gamma-ray emission is negligible. 
In the \textit{pp}-dominated scenario, $\pi^{0/\pm}$ decays arising from \textit{pp} interactions in the SAM dominate the GeV gamma-ray/neutrino emission. The EC emission from the SAM is larger than that from the SW because the SAM has more secondary electrons.
In the EC-dominated scenario, the EC emission from the SAM and the SW region make a comparable contribution to the total GeV gamma-ray flux. The expected neutrino flux is $\sim10$ times smaller than that of \textit{pp}-dominated scenario because of the lower ISM density.

\section{Discussion}\label{sec:Discussion}
\subsection{Distinguishing Between the Emission Scenarios}
\begin{figure}
    \centering
    \includegraphics[width=1\linewidth]{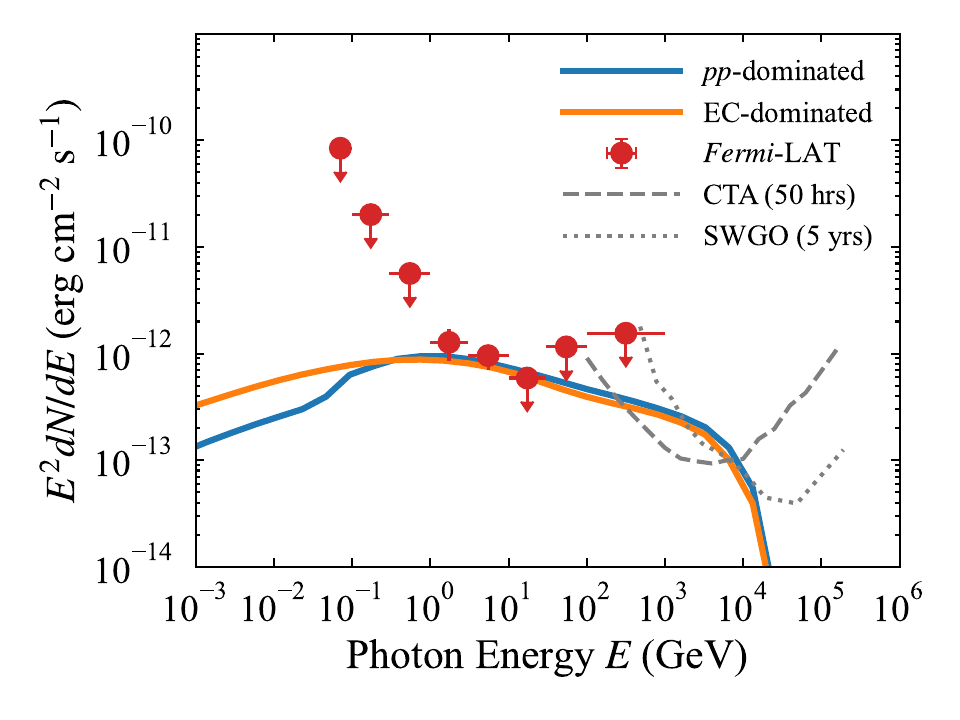}
    \caption{Same as Figure \ref{fig:sed}, but showing the energy range of $10^{-3}\mathrm{-}10^{6}~\mathrm{GeV}$. The blue and orange lines show the gamma-ray spectra for the $pp$-dominated and EC-dominated scenarios. Red circles show \textit{Fermi}-LAT data \citep{Ballet+2023}. Gray dashed and dotted lines show the sensitivity curves of CTA with a 50-hour integration time \citep{CTA+2019}, and SWGO after 5 years of observations \citep{Huentemeyer+2019}.}
    \label{fig:sed_gamma}
\end{figure}

Figure~\ref{fig:sed_gamma} provides a detailed view in energy range 1~MeV--1~PeV. Future gamma-ray observations with instruments such as the Cherenkov Telescope Array \citep[CTA;][]{CTA+2019}\footnote{\url{https://www.ctao.org/for-scientists/performance/}} and the Southern Wide-field Gamma-ray Observatory \citep[SWGO;][]{Huentemeyer+2019}\footnote{\url{https://www.swgo.org/SWGOWiki/doku.php}} could detect GRS~1732-292 at TeV energies, and would have the potential to improve constraints on the CR acceleration efficiency in this system, $\eta_\mathrm{g}$, which is important when estimating the maximum energy to which GRS~1734-292 can accelerate CRs. 
However, both the \textit{pp}-dominated and EC-dominated scenarios yield similar spectral shapes in the GeV--TeV bands, making discrimination between them challenging with current and near-future gamma-ray observations (Figure~\ref{fig:sed_gamma}). While future MeV missions like COSI \citep{Zoglauer+2021}\footnote{\url{https://cosi.ssl.berkeley.edu/}} and GRAMS \citep{Aramaki+20}\footnote{\url{https://grams.sites.northeastern.edu/}} might detect a characteristic pion-decay cutoff at  $\lesssim70$~MeV in the \textit{pp}-dominated scenario, potential contamination from coronal emission up to $\lesssim1$~GeV \citep{Inoue:2019fil} could complicate this distinction.

Neutrino observations with next-generation facilities like IceCube-Gen2 \citep{Aartsen+2021}, KM3Net \citep{Aiello+2024}\footnote{\url{https://www.km3net.org/}}, TRIDENT \citep{Ye+2023}\footnote{\url{https://trident.sjtu.edu.cn/en}}, and P-ONE \citep{Malecki2024}\footnote{\url{https://www.pacific-neutrino.org/}} offer a promising path to distinguish between these scenarios. These observations could definitively discriminate between the \textit{pp}-dominated and EC-dominated cases if sensitivities of $\lesssim10^{-10}\ \mathrm{GeV\ cm^{-2}\ s^{-1}}$ at $1\ \mathrm{TeV}$ are achieved (see Figure \ref{fig:sed}).

Additionally, ALMA observations of the central $100\ \mathrm{pc}$ region could test the \textit{pp}-dominated scenario, as it requires an abundant gas target. Given the required gas density (see Table~\ref{tab:fiducial parameter}), the predicted CO ($J=1\rightarrow0$) flux density of $\sim32\ \mathrm{mJy}$, based on the empirical conversion \citep{2013ARA&A..51..207B} adopting a typical CO abundance of $\mathrm{H_2:CO=5000:1}$ \citep{Lacy+1994}, would be readily detectable\footnote{An integration time of a few minutes would be required for a clear detection, see
\url{https://almascience.nao.ac.jp/proposing/sensitivity-calculator}.}. This amount of gas still remains consistent with far-infrared constraints (see Section~\ref{sec: GRS}).

\subsection{Comparison with NGC~1068 and NGC~4151}
Recent observations have established that some Seyfert galaxies are sources of both high-energy gamma-rays and neutrinos. Currently, two such objects have been detected: NGC~1068 \citep{IceCube+22} and NGC~4151 \citep{IceCube2024arXiv240606684A, IceCube2024arXiv240607601A}\footnote{While there are two gamma-ray bright blazars within the IceCube error circle of NGC~4151 \citep{Buson:2023irp, Murase2024ApJ...961L..34M}, these sources would not significantly contribute to the IceCube signals \citep[e.g.,][]{Omeliukh:2024fhg}.}. Here, we compare GRS~1734-292 with these objects.

The origin of gamma-ray emission differs among these objects. Like GRS~1734-292, pc-scale disk wind models could explain the gamma-ray emission from NGC~1068 \citep{Lamastra+16, Peretti+23} and NGC~4151  \citep{Peretti:2023crf}. However, their gamma-ray emission could also be readily explained either by star-formation activity \citep{Eichmann:2022lxh, Ajello:2023hkh} or weak jets \citep{Inoue:2023bmy, Fang+2024}. In contrast, GRS~1734-292 stands out because, as demonstrated in Section~\ref{sec: GRS}, its star-formation and jet activities\footnote{There is also a warm absorber component in GRS 1734-292 \citep{Tortosa+17} but its kinetic power \citep[$<2\times10^{44}\ \mathrm{erg\ s^{-1}}$, estimated from][]{Blustin+2005} cannot exceed even 10\% of the UFO kinetic power. We therefore neglect its contribution to the gamma-ray emission.} are too weak to account for the observed gamma-ray emission, making the disk wind scenario particularly compelling. This distinctive feature makes GRS~1734-292 a unique laboratory for testing the disk wind scenario.

The detected neutrino emission from NGC~1068 and NGC~4151 likely originates in accreting coronae near their supermassive black holes  \citep[e.g.,][]{Inoue2020ApJ...891L..33I, Murase2024ApJ...961L..34M}. This coronal origin is favored because the neutrino-emitting regions must be gamma-ray opaque to explain the observed gamma-ray deficit \citep[but see also][]{Hooper:2023ssc,Inoue:2023bmy,Fang+2024,Yasuda:2024fvc,Inoue:2024nap}. While large-scale disk wind models like the one we apply to GRS~1734-292 cannot reproduce these neutrino signals in NGC~1068 and NGC~4151 \citep{Peretti+23, Peretti:2023crf}, coronal-scale winds remain viable \citep{Inoue+22, Huang:2024yua}. Although no neutrino emission has yet been reported from GRS~1734-292, future detection of a neutrino flux exceeding the gamma-ray flux would suggest a coronal origin similar to NGC~1068 and NGC~4151.

\subsection{Implications for Particle Acceleration in AGN Disk Winds}\label{sec:UHECR}

Our results demonstrate that AGN disk winds in Seyfert galaxies could be significant particle accelerators. In both scenarios, the reverse shock of GRS 1734-292 could accelerate protons to energies of $\approx2\times10^{18}$~eV and $\approx9\times10^{17}$~eV for the \textit{pp}-dominated and EC-dominated scenarios, respectively (see Figure \ref{fig:timescales_pp} and \ref{fig:timescales_EC}), with even higher energies possible for heavier nuclei. This supports the emerging picture of Seyfert galaxies as potential ultra-high energy cosmic ray (UHECR) sources \citep[see][]{Peretti+23, 2024arXiv241105667E}.

\section{Conclusions}\label{sec:Conclusions}

We have investigated the origin of GeV gamma-ray emission from the nearby Seyfert galaxy GRS 1734-292, which presents an intriguing case where both starburst and jet activities are too weak to explain the observed gamma-ray flux. This distinguishes it from other gamma-ray detected Seyfert galaxies like NGC 1068 and NGC 4151, where the GeV emission can be attributed to starburst or jet activity.

Using a detailed AGN disk wind model that accounts for wind-ISM interactions and the resulting shocked regions, we demonstrate that disk winds can explain the observed GeV gamma-ray emission. Our analysis reveals two viable scenarios that can reproduce the \textit{Fermi}-LAT data: a hadronic $pp$-dominated scenario requiring dense ISM ($n_0=200~\mathrm{cm^{-3}}$) and evolved system age ($t_\mathrm{wind}=10^5~\mathrm{yr}$), and a leptonic EC-dominated scenario with more moderate ISM density ($n_0=10~\mathrm{cm^{-3}}$) and younger age ($t_\mathrm{wind}=3\times10^3~\mathrm{yr}$). These results establish GRS 1734-292 as an important laboratory for testing AGN disk wind models and their role in particle acceleration. Future multi-messenger observations with TeV neutrino detectors and ALMA will be crucial in discriminating between these scenarios and advancing our understanding of particle acceleration mechanisms in radio-quiet AGN.

\section*{Acknowledgements}
The authors thank the anonymous referee for their helpful comments which improved the manuscript. The authors also thank Samuel Barnier, Susumu Inoue, and Hirokazu Odaka for useful comments and discussions.
This research has made use of the NASA/IPAC Extragalactic Database (NED), which is funded by the National Aeronautics and Space Administration and operated by the California Institute of Technology. YI is supported by NAOJ ALMA Scientific Research Grant Number 2021-17A; JSPS KAKENHI Grant Number JP18H05458, JP19K14772, and JP22K18277; and World Premier International Research Center Initiative (WPI), MEXT, Japan. ERO is a JSPS international research fellow, supported by JSPS KAKENHI Grant Number JP22F22327. TM was supported (in part) by a University Research Support Grant from the National Astronomical Observatory of Japan (NAOJ). RT acknowledges a Grant-in-Aid for JSPS Fellows, Grant Number JP24KJ0152.

\bibliography{ref}{}

\begin{thebibliography}{}
\expandafter\ifx\csname natexlab\endcsname\relax\def\natexlab#1{#1}\fi
\providecommand{\url}[1]{\href{#1}{#1}}
\providecommand{\dodoi}[1]{doi:~\href{http://doi.org/#1}{\nolinkurl{#1}}}
\providecommand{\doeprint}[1]{\href{http://ascl.net/#1}{\nolinkurl{http://ascl.net/#1}}}
\providecommand{\doarXiv}[1]{\href{https://arxiv.org/abs/#1}{\nolinkurl{https://arxiv.org/abs/#1}}}

\bibitem[{{Aartsen} {et~al.}(2021){Aartsen}, {Abbasi}, {Ackermann}, {Adams}, {Aguilar}, {Ahlers}, {Ahrens}, {Alispach}, {Allison}, {Amin}, {Andeen}, {Anderson}, {Ansseau}, {Anton}, {Arg{\"u}elles}, {Arlen}, {Auffenberg}, {Axani}, {Bagherpour}, {Bai}, {Balagopal V}, {Barbano}, {Bartos}, {Bastian}, {Basu}, {Baum}, {Baur}, {Bay}, {Beatty}, {Becker}, {Tjus}, {BenZvi}, {Berley}, {Bernardini}, {Besson}, {Binder}, {Bindig}, {Blaufuss}, {Blot}, {Bohm}, {Bohmer}, {B{\"o}ser}, {Botner}, {B{\"o}ttcher}, {Bourbeau}, {Bourbeau}, {Bradascio}, {Braun}, {Bron}, {Brostean-Kaiser}, {Burgman}, {Burley}, {Buscher}, {Busse}, {Bustamante}, {Campana}, {Carnie-Bronca}, {Carver}, {Chen}, {Chen}, {Cheung}, {Chirkin}, {Choi}, {Clark}, {Clark}, {Classen}, {Coleman}, {Collin}, {Connolly}, {Conrad}, {Coppin}, {Correa}, {Cowen}, {Cross}, {Dave}, {Deaconu}, {De Clercq}, {DeLaunay}, {De Kockere}, {Dembinski}, {Deoskar}, {De Ridder}, {Desai}, {Desiati}, {de Vries}, {de Wasseige}, {de With}, {DeYoung}, {Dharani}, {Diaz}, {D{\'\i}az-V{\'e}lez},
  {Dujmovic}, {Dunkman}, {DuVernois}, {Dvorak}, {Ehrhardt}, {Eller}, {Engel}, {Evans}, {Evenson}, {Fahey}, {Farrag}, {Fazely}, {Felde}, {Fienberg}, {Filimonov}, {Finley}, {Fischer}, {Fox}, {Franckowiak}, {Friedman}, {Fritz}, {Gaisser}, {Gallagher}, {Ganster}, {Garcia-Fernandez}, {Garrappa}, {Gartner}, {Gerhard}, {Gernhaeuser}, {Ghadimi}, {Glaser}, {Glauch}, {Gl{\"u}senkamp}, {Goldschmidt}, {Gonzalez}, {Goswami}, {Grant}, {Gr{\'e}goire}, {Griffith}, {Griswold}, {G{\"u}nd{\"u}z}, {Haack}, {Hallgren}, {Halliday}, {Halve}, {Halzen}, {Hanson}, {Hanson}, {Hardin}, {Haugen}, {Haungs}, {Hauser}, {Hebecker}, {Heinen}, {Heix}, {Helbing}, {Hellauer}, {Henningsen}, {Hickford}, {Hignight}, {Hill}, {Hill}, {Hoffman}, {Hoffmann}, {Hoffmann}, {Hoinka}, {Hokanson-Fasig}, {Holzapfel}, {Hoshina}, {Huang}, {Huber}, {Huber}, {Huege}, {Hughes}, {Hultqvist}, {H{\"u}nnefeld}, {Hussain}, {In}, {Iovine}, {Ishihara}, {Jansson}, {Japaridze}, {Jeong}, {Jones}, {Jonske}, {Joppe}, {Kalekin}, {Kang}, {Kang}, {Kang}, {Kappes}, {Kappesser},
  {Karg}, {Karl}, {Karle}, {Katori}, {Katz}, {Kauer}, {Keivani}, {Kellermann}, {Kelley}, {Kheirandish}, {Kim}, {Kin}, {Kintscher}, {Kiryluk}, {Kittler}, {Kleifges}, {Klein}, {Koirala}, {Kolanoski}, {K{\"o}pke}, {Kopper}, {Kopper}, {Koskinen}, {Koundal}, {Kovacevich}, {Kowalski}, {Krauss}, {Krings}, {Kr{\"u}ckl}, {Kulacz}, {Kurahashi}, {Gualda}, {Lahmann}, {Lanfranchi}, {Larson}, {Latif}, {Lauber}, {Lazar}, {Leonard}, {Leszczy{\'n}ska}, {Li}, {Liu}, {Lohfink}, {LoSecco}, {Mariscal}, {Lu}, {Lucarelli}, {Ludwig}, {L{\"u}nemann}, {Luszczak}, {Lyu}, {Ma}, {Madsen}, {Maggi}, {Mahn}, {Makino}, {Mallik}, {Mancina}, {Mandalia}, {Mari{\c{s}}}, {Marka}, {Marka}, {Maruyama}, {Mase}, {Maunu}, {McNally}, {Meagher}, {Medina}, {Meier}, {Meighen-Berger}, {Merz}, {Meyers}, {Micallef}, {Mockler}, {Moment{\'e}}, {Montaruli}, {Moore}, {Morse}, {Moulai}, {Muth}, {Naab}, {Nagai}, {Nam}, {Nauman}, {Necker}, {Neer}, {Nelles}, {Nguyễn}, {Niederhausen}, {Nisa}, {Nowicki}, {Nygren}, {Oberla}, {Pollmann}, {Oehler}, {Olivas},
  {O'Sullivan}, {Pan}, {Pandya}, {Pankova}, {Papp}, {Park}, {Parker}, {Paudel}, {Peiffer}, {P{\'e}rez de los Heros}, {Petersen}, {Philippen}, {Pieloth}, {Pieper}, {Pinfold}, {Pizzuto}, {Plaisier}, {Plum}, {Popovych}, {Porcelli}, {Rodriguez}, {Price}, {Przybylski}, {Raab}, {Raissi}, {Rameez}, {Rauch}, {Rawlins}, {Rea}, {Rehman}, {Reimann}, {Renschler}, {Renzi}, {Resconi}, {Reusch}, {Rhode}, {Richman}, {Riedel}, {Riegel}, {Roberts}, {Robertson}, {Roellinghoff}, {Rongen}, {Rott}, {Ruhe}, {Ryckbosch}, {Cantu}, {Safa}, {Herrera}, {Sandrock}, {Sandroos}, {Sandstrom}, {Santander}, {Sarkar}, {Sarkar}, {Satalecka}, {Scharf}, {Schaufel}, {Schieler}, {Schlunder}, {Schmidt}, {Schneider}, {Schneider}, {Schr{\"o}der}, {Schumacher}, {Sclafani}, {Seckel}, {Seunarine}, {Shaevitz}, {Sharma}, {Shefali}, {Silva}, {Smith}, {Smithers}, {Snihur}, {Soedingrekso}, {Soldin}, {S{\"o}ldner-Rembold}, {Song}, {Southall}, {Spiczak}, {Spiering}, {Stachurska}, {Stamatikos}, {Stanev}, {Stein}, {Stettner}, {Steuer}, {Stezelberger}, {Stokstad},
  {Strotjohann}, {St{\"u}rwald}, {Stuttard}, {Sullivan}, {Taboada}, {Taketa}, {Tanaka}, {Tenholt}, {Ter-Antonyan}, {Terliuk}, {Tilav}, {Tollefson}, {Tomankova}, {T{\"o}nnis}, {Torres}, {Toscano}, {Tosi}, {Trettin}, {Tselengidou}, {Tung}, {Turcati}, {Turcotte}, {Turley}, {Twagirayezu}, {Ty}, {Unger}, {Elorrieta}, {Vandenbroucke}, {van Eijk}, {van Eijndhoven}, {Vannerom}, {van Santen}, {Veberic}, {Verpoest}, {Vieregg}, {Vraeghe}, {Walck}, {Watson}, {Weaver}, {Weindl}, {Weinstock}, {Weiss}, {Weldert}, {Welling}, {Wendt}, {Werthebach}, {Whitehorn}, {Wiebe}, {Wiebusch}, {Williams}, {Wissel}, {Wolf}, {Wood}, {Woschnagg}, {Wrede}, {Wren}, {Wulff}, {Xu}, {Xu}, {Yanez}, {Yoshida}, {Yuan}, {Zhang}, {Zierke}, \& {Z{\"o}cklein}}]{Aartsen+2021}
{Aartsen}, M.~G., {Abbasi}, R., {Ackermann}, M., {et~al.} 2021, Journal of Physics G Nuclear Physics, 48, 060501, \dodoi{10.1088/1361-6471/abbd48}

\bibitem[{{Abbasi} {et~al.}(2024{\natexlab{a}}){Abbasi}, {Ackermann}, {Adams}, {Agarwalla}, {Aguilar}, {Ahlers}, {Alameddine}, {Amin}, {Andeen}, {Arg{\"u}elles}, {Ashida}, {Athanasiadou}, {Ausborm}, {Axani}, {Bai}, {Balagopal V.}, {Baricevic}, {Barwick}, {Bash}, {Basu}, {Bay}, {Beatty}, {Becker Tjus}, {Beise}, {Bellenghi}, {Benning}, {BenZvi}, {Berley}, {Bernardini}, {Besson}, {Blaufuss}, {Bloom}, {Blot}, {Bontempo}, {Book Motzkin}, {Boscolo Meneguolo}, {B{\"o}ser}, {Botner}, {B{\"o}ttcher}, {Braun}, {Brinson}, {Brostean-Kaiser}, {Brusa}, {Burley}, {Butterfield}, {Campana}, {Caracas}, {Carloni}, {Carpio}, {Chattopadhyay}, {Chau}, {Chen}, {Chirkin}, {Choi}, {Clark}, {Coleman}, {Collin}, {Connolly}, {Conrad}, {Coppin}, {Corley}, {Correa}, {Cowen}, {Dave}, {De Clercq}, {DeLaunay}, {Delgado}, {Deng}, {Desai}, {Desiati}, {de Vries}, {de Wasseige}, {DeYoung}, {Diaz}, {D{\'\i}az-V{\'e}lez}, {Dierichs}, {Dittmer}, {Domi}, {Draper}, {Dujmovic}, {Dutta}, {DuVernois}, {Ehrhardt}, {Eidenschink}, {Eimer}, {Eller},
  {Ellinger}, {El Mentawi}, {Els{\"a}sser}, {Engel}, {Erpenbeck}, {Evans}, {Evenson}, {Fan}, {Fang}, {Farrag}, {Fazely}, {Fedynitch}, {Feigl}, {Fiedlschuster}, {Finley}, {Fischer}, {Fox}, {Franckowiak}, {Fukami}, {F{\"u}rst}, {Gallagher}, {Ganster}, {Garcia}, {Garcia}, {Garg}, {Genton}, {Gerhardt}, {Ghadimi}, {Girard-Carillo}, {Glaser}, {Gl{\"u}senkamp}, {Gonzalez}, {Goswami}, {Granados}, {Grant}, {Gray}, {Gries}, {Griffin}, {Griswold}, {Groth}, {G{\"u}nther}, {Gutjahr}, {Ha}, {Haack}, {Hallgren}, {Halve}, {Halzen}, {Hamdaoui}, {Minh}, {Handt}, {Hanson}, {Hardin}, {Harnisch}, {Hatch}, {Haungs}, {H{\"a}u{\ss}ler}, {Helbing}, {Hellrung}, {Hermannsgabner}, {Heuermann}, {Heyer}, {Hickford}, {Hidvegi}, {Hill}, {Hill}, {Hoffman}, {Hori}, {Hoshina}, {Hostert}, {Hou}, {Huber}, {Hultqvist}, {H{\"u}nnefeld}, {Hussain}, {Hymon}, {Ishihara}, {Iwakiri}, {Jacquart}, {Jain}, {Janik}, {Jansson}, {Japaridze}, {Jeong}, {Jin}, {Jones}, {Kamp}, {Kang}, {Kang}, {Kang}, {Kappes}, {Kappesser}, {Kardum}, {Karg}, {Karl}, {Karle},
  {Katil}, {Katz}, {Kauer}, {Kelley}, {Khanal}, {Khatee Zathul}, {Kheirandish}, {Kiryluk}, {Klein}, {Kochocki}, {Koirala}, {Kolanoski}, {Kontrimas}, {K{\"o}pke}, {Kopper}, {Koskinen}, {Koundal}, {Kovacevich}, {Kowalski}, {Kozynets}, {Krishnamoorthi}, {Kruiswijk}, {Krupczak}, {Kumar}, {Kun}, {Kurahashi}, {Lad}, {Lagunas Gualda}, {Lamoureux}, {Larson}, {Latseva}, {Lauber}, {Lazar}, {Lee}, {DeHolton}, {Leszczy{\'n}ska}, {Liao}, {Lincetto}, {Liu}, {Liubarska}, {Love}, {Lozano Mariscal}, {Lu}, {Lucarelli}, {Luszczak}, {Lyu}, {Madsen}, {Magnus}, {Mahn}, {Makino}, {Manao}, {Mancina}, {Sainte}, {Mari{\c{s}}}, {Marka}, {Marka}, {Marsee}, {Martinez-Soler}, {Maruyama}, {Mayhew}, {McNally}, {Mead}, {Meagher}, {Mechbal}, {Medina}, {Meier}, {Merckx}, {Merten}, {Micallef}, {Mitchell}, {Montaruli}, {Moore}, {Morii}, {Morse}, {Moulai}, {Mukherjee}, {Naab}, {Nagai}, {Nakos}, {Naumann}, {Necker}, {Negi}, {Neste}, {Neumann}, {Niederhausen}, {Nisa}, {Noda}, {Noell}, {Novikov}, {Obertacke Pollmann}, {O'Dell}, {Oeyen}, {Olivas},
  {Orsoe}, {Osborn}, {O'Sullivan}, {Palusova}, {Pandya}, {Park}, {Parker}, {Paudel}, {Paul}, {P{\'e}rez de los Heros}, {Pernice}, {Peterson}, {Philippen}, {Pizzuto}, {Plum}, {Pont{\'e}n}, {Popovych}, {Prado Rodriguez}, {Pries}, {Privon}, {Procter-Murphy}, {Przybylski}, {Raab}, {Rack-Helleis}, {Ravn}, {Rawlins}, {Rechav}, {Rehman}, {Reichherzer}, {Resconi}, {Reusch}, {Rhode}, {Riedel}, {Rifaie}, {Roberts}, {Robertson}, {Rodan}, {Roellinghoff}, {Rongen}, {Rosted}, {Rott}, {Ruhe}, {Ruohan}, {Ryckbosch}, {Safa}, {Saffer}, {Salazar-Gallegos}, {Sampathkumar}, {Sandrock}, {Santander}, {Sarkar}, {Sarkar}, {Savelberg}, {Savina}, {Schaile}, {Schaufel}, {Schieler}, {Schindler}, {Schlickmann}, {Schl{\"u}ter}, {Schl{\"u}ter}, {Schmeisser}, {Schmidt}, {Schneider}, {Schr{\"o}der}, {Schumacher}, {Sclafani}, {Seckel}, {Seikh}, {Seo}, {Seunarine}, {Sevle Myhr}, {Shah}, {Shefali}, {Shimizu}, {Silva}, {Skrzypek}, {Smithers}, {Snihur}, {Soedingrekso}, {S{\o}gaard}, {Soldin}, {Soldin}, {Sommani}, {Spannfellner}, {Spiczak},
  {Spiering}, {Stamatikos}, {Stanev}, {Stezelberger}, {St{\"u}rwald}, {Stuttard}, {Sullivan}, {Taboada}, {Ter-Antonyan}, {Terliuk}, {Thiesmeyer}, {Thompson}, {Thwaites}, {Tilav}, {Tollefson}, {T{\"o}nnis}, {Toscano}, {Tosi}, {Trettin}, {Turcotte}, {Twagirayezu}, {Unland Elorrieta}, {Upadhyay}, {Upshaw}, {Vaidyanathan}, {Valtonen-Mattila}, {Vandenbroucke}, {van Eijndhoven}, {Vannerom}, {van Santen}, {Vara}, {Varsi}, {Veitch-Michaelis}, {Venugopal}, {Vereecken}, {Verpoest}, {Veske}, {Vijai}, {Walck}, {Wang}, {Weaver}, {Weigel}, {Weindl}, {Weldert}, {Wen}, {Wendt}, {Werthebach}, {Weyrauch}, {Whitehorn}, {Wiebusch}, {Williams}, {Witthaus}, {Wolf}, {Wolf}, {Wrede}, {Xu}, {Yanez}, {Yildizci}, {Yoshida}, {Young}, {Yu}, {Yuan}, {Zhang}, {Zhelnin}, {Zilberman}, \& {Zimmerman}}]{IceCube2024arXiv240606684A}
{Abbasi}, R., {Ackermann}, M., {Adams}, J., {et~al.} 2024{\natexlab{a}}, arXiv e-prints, arXiv:2406.06684, \dodoi{10.48550/arXiv.2406.06684}

\bibitem[{{Abbasi} {et~al.}(2024{\natexlab{b}}){Abbasi}, {Ackermann}, {Adams}, {Agarwalla}, {Aguilar}, {Ahlers}, {Alameddine}, {Amin}, {Andeen}, {Arg{\"u}elles}, {Ashida}, {Athanasiadou}, {Ausborm}, {Axani}, {Bai}, {Balagopal V.}, {Baricevic}, {Barwick}, {Bash}, {Basu}, {Bay}, {Beatty}, {Becker Tjus}, {Beise}, {Bellenghi}, {Benning}, {BenZvi}, {Berley}, {Bernardini}, {Besson}, {Blaufuss}, {Bloom}, {Blot}, {Bontempo}, {Book Motzkin}, {Boscolo Meneguolo}, {B{\"o}ser}, {Botner}, {B{\"o}ttcher}, {Braun}, {Brinson}, {Brostean-Kaiser}, {Brusa}, {Burley}, {Butterfield}, {Campana}, {Caracas}, {Carloni}, {Carpio}, {Chattopadhyay}, {Chau}, {Chen}, {Chirkin}, {Choi}, {Clark}, {Coleman}, {Collin}, {Connolly}, {Conrad}, {Coppin}, {Corley}, {Correa}, {Cowen}, {Dave}, {De Clercq}, {DeLaunay}, {Delgado}, {Deng}, {Desai}, {Desiati}, {de Vries}, {de Wasseige}, {DeYoung}, {Diaz}, {D{\'\i}az-V{\'e}lez}, {Dierichs}, {Dittmer}, {Domi}, {Draper}, {Dujmovic}, {Dutta}, {DuVernois}, {Ehrhardt}, {Eidenschink}, {Eimer}, {Eller},
  {Ellinger}, {El Mentawi}, {Els{\"a}sser}, {Engel}, {Erpenbeck}, {Evans}, {Evenson}, {Fan}, {Fang}, {Farrag}, {Fazely}, {Fedynitch}, {Feigl}, {Fiedlschuster}, {Finley}, {Fischer}, {Fox}, {Franckowiak}, {Fukami}, {F{\"u}rst}, {Gallagher}, {Ganster}, {Garcia}, {Garcia}, {Garg}, {Genton}, {Gerhardt}, {Ghadimi}, {Girard-Carillo}, {Glaser}, {Glauch}, {Gl{\"u}senkamp}, {Gonzalez}, {Goswami}, {Granados}, {Grant}, {Gray}, {Gries}, {Griffin}, {Griswold}, {Groth}, {G{\"u}nther}, {Gutjahr}, {Ha}, {Haack}, {Hallgren}, {Halve}, {Halzen}, {Hamdaoui}, {Minh}, {Handt}, {Hanson}, {Hardin}, {Harnisch}, {Hatch}, {Haungs}, {H{\"a}u{\ss}ler}, {Helbing}, {Hellrung}, {Hermannsgabner}, {Heuermann}, {Heyer}, {Hickford}, {Hidvegi}, {Hill}, {Hill}, {Hoffman}, {Hori}, {Hoshina}, {Hostert}, {Hou}, {Huber}, {Hultqvist}, {H{\"u}nnefeld}, {Hussain}, {Hymon}, {Ishihara}, {Iwakiri}, {Jacquart}, {Janik}, {Jansson}, {Japaridze}, {Jeong}, {Jin}, {Jones}, {Kamp}, {Kang}, {Kang}, {Kang}, {Kappes}, {Kappesser}, {Kardum}, {Karg}, {Karl}, {Karle},
  {Katil}, {Katz}, {Kauer}, {Kelley}, {Khanal}, {Khatee Zathul}, {Kheirandish}, {Kiryluk}, {Klein}, {Kochocki}, {Koirala}, {Kolanoski}, {Kontrimas}, {K{\"o}pke}, {Kopper}, {Koskinen}, {Koundal}, {Kovacevich}, {Kowalski}, {Kozynets}, {Krishnamoorthi}, {Kruiswijk}, {Krupczak}, {Kumar}, {Kun}, {Kurahashi}, {Lad}, {Lagunas Gualda}, {Lamoureux}, {Larson}, {Latseva}, {Lauber}, {Lazar}, {Lee}, {DeHolton}, {Leszczy{\'n}ska}, {Liao}, {Lincetto}, {Liu}, {Liu}, {Liubarska}, {Lohfink}, {Love}, {Lozano Mariscal}, {Lu}, {Lucarelli}, {Luszczak}, {Lyu}, {Madsen}, {Magnus}, {Mahn}, {Makino}, {Manao}, {Mancina}, {Sainte}, {Mari{\c{s}}}, {Marka}, {Marka}, {Marsee}, {Martinez-Soler}, {Maruyama}, {Mayhew}, {McNally}, {Mead}, {Meagher}, {Mechbal}, {Medina}, {Meier}, {Merckx}, {Merten}, {Micallef}, {Mitchell}, {Montaruli}, {Moore}, {Morii}, {Morse}, {Moulai}, {Mukherjee}, {Naab}, {Nagai}, {Nakos}, {Naumann}, {Necker}, {Negi}, {Neste}, {Neumann}, {Niederhausen}, {Nisa}, {Noda}, {Noell}, {Novikov}, {Obertacke Pollmann}, {O'Dell},
  {Oeyen}, {Olivas}, {Orsoe}, {Osborn}, {O'Sullivan}, {Pandya}, {Park}, {Parker}, {Paudel}, {Paul}, {P{\'e}rez de los Heros}, {Pernice}, {Peterson}, {Philippen}, {Pizzuto}, {Plum}, {Pont{\'e}n}, {Popovych}, {Prado Rodriguez}, {Pries}, {Procter-Murphy}, {Przybylski}, {Raab}, {Rack-Helleis}, {Ravn}, {Rawlins}, {Rechav}, {Rehman}, {Reichherzer}, {Resconi}, {Reusch}, {Rhode}, {Riedel}, {Rifaie}, {Roberts}, {Robertson}, {Rodan}, {Roellinghoff}, {Rongen}, {Rosted}, {Rott}, {Ruhe}, {Ruohan}, {Ryckbosch}, {Safa}, {Saffer}, {Salazar-Gallegos}, {Sampathkumar}, {Sandrock}, {Santander}, {Sarkar}, {Sarkar}, {Savelberg}, {Savina}, {Schaile}, {Schaufel}, {Schieler}, {Schindler}, {Schl{\"u}ter}, {Schl{\"u}ter}, {Schmeisser}, {Schmidt}, {Schneider}, {Schr{\"o}der}, {Schumacher}, {Sclafani}, {Seckel}, {Seikh}, {Seo}, {Seunarine}, {Sevle Myhr}, {Shah}, {Shefali}, {Shimizu}, {Silva}, {Skrzypek}, {Smithers}, {Snihur}, {Soedingrekso}, {S{\o}gaard}, {Soldin}, {Soldin}, {Sommani}, {Spannfellner}, {Spiczak}, {Spiering}, {Stamatikos},
  {Stanev}, {Stezelberger}, {St{\"u}rwald}, {Stuttard}, {Sullivan}, {Taboada}, {Ter-Antonyan}, {Terliuk}, {Thiesmeyer}, {Thompson}, {Thwaites}, {Tilav}, {Tollefson}, {T{\"o}nnis}, {Toscano}, {Tosi}, {Trettin}, {Turcotte}, {Twagirayezu}, {Unland Elorrieta}, {Upadhyay}, {Upshaw}, {Vaidyanathan}, {Valtonen-Mattila}, {Vandenbroucke}, {van Eijndhoven}, {Vannerom}, {van Santen}, {Vara}, {Varsi}, {Veitch-Michaelis}, {Venugopal}, {Vereecken}, {Verpoest}, {Veske}, {Vijai}, {Walck}, {Wang}, {Weaver}, {Weigel}, {Weindl}, {Weldert}, {Wen}, {Wendt}, {Werthebach}, {Weyrauch}, {Whitehorn}, {Wiebusch}, {Williams}, {Witthaus}, {Wolf}, {Wolf}, {Wrede}, {Xu}, {Yanez}, {Yildizci}, {Yoshida}, {Young}, {Yu}, {Yuan}, {Zhang}, {Zhelnin}, {Zilberman}, \& {Zimmerman}}]{IceCube2024arXiv240607601A}
---. 2024{\natexlab{b}}, arXiv e-prints, arXiv:2406.07601, \dodoi{10.48550/arXiv.2406.07601}

\bibitem[{{Abdollahi} {et~al.}(2020){Abdollahi}, {Acero}, {Ackermann}, {Ajello}, {Atwood}, {Axelsson}, {Baldini}, {Ballet}, {Barbiellini}, {Bastieri}, {Becerra Gonzalez}, {Bellazzini}, {Berretta}, {Bissaldi}, {Blandford}, {Bloom}, {Bonino}, {Bottacini}, {Brandt}, {Bregeon}, {Bruel}, {Buehler}, {Burnett}, {Buson}, {Cameron}, {Caputo}, {Caraveo}, {Casandjian}, {Castro}, {Cavazzuti}, {Charles}, {Chaty}, {Chen}, {Cheung}, {Chiaro}, {Ciprini}, {Cohen-Tanugi}, {Cominsky}, {Coronado-Bl{\'a}zquez}, {Costantin}, {Cuoco}, {Cutini}, {D'Ammando}, {DeKlotz}, {de la Torre Luque}, {de Palma}, {Desai}, {Digel}, {Di Lalla}, {Di Mauro}, {Di Venere}, {Dom{\'\i}nguez}, {Dumora}, {Fana Dirirsa}, {Fegan}, {Ferrara}, {Franckowiak}, {Fukazawa}, {Funk}, {Fusco}, {Gargano}, {Gasparrini}, {Giglietto}, {Giommi}, {Giordano}, {Giroletti}, {Glanzman}, {Green}, {Grenier}, {Griffin}, {Grondin}, {Grove}, {Guiriec}, {Harding}, {Hayashi}, {Hays}, {Hewitt}, {Horan}, {J{\'o}hannesson}, {Johnson}, {Kamae}, {Kerr}, {Kocevski}, {Kovac'evic'},
  {Kuss}, {Landriu}, {Larsson}, {Latronico}, {Lemoine-Goumard}, {Li}, {Liodakis}, {Longo}, {Loparco}, {Lott}, {Lovellette}, {Lubrano}, {Madejski}, {Maldera}, {Malyshev}, {Manfreda}, {Marchesini}, {Marcotulli}, {Mart{\'\i}-Devesa}, {Martin}, {Massaro}, {Mazziotta}, {McEnery}, {Mereu}, {Meyer}, {Michelson}, {Mirabal}, {Mizuno}, {Monzani}, {Morselli}, {Moskalenko}, {Negro}, {Nuss}, {Ojha}, {Omodei}, {Orienti}, {Orlando}, {Ormes}, {Palatiello}, {Paliya}, {Paneque}, {Pei}, {Pe{\~n}a-Herazo}, {Perkins}, {Persic}, {Pesce-Rollins}, {Petrosian}, {Petrov}, {Piron}, {Poon}, {Porter}, {Principe}, {Rain{\`o}}, {Rando}, {Razzano}, {Razzaque}, {Reimer}, {Reimer}, {Remy}, {Reposeur}, {Romani}, {Saz Parkinson}, {Schinzel}, {Serini}, {Sgr{\`o}}, {Siskind}, {Smith}, {Spandre}, {Spinelli}, {Strong}, {Suson}, {Tajima}, {Takahashi}, {Tak}, {Thayer}, {Thompson}, {Tibaldo}, {Torres}, {Torresi}, {Valverde}, {Van Klaveren}, {van Zyl}, {Wood}, {Yassine}, \& {Zaharijas}}]{Abdollahi+20}
{Abdollahi}, S., {Acero}, F., {Ackermann}, M., {et~al.} 2020, \apjs, 247, 33, \dodoi{10.3847/1538-4365/ab6bcb}

\bibitem[{{Abdollahi} {et~al.}(2022){Abdollahi}, {Acero}, {Baldini}, {Ballet}, {Bastieri}, {Bellazzini}, {Berenji}, {Berretta}, {Bissaldi}, {Blandford}, {Bloom}, {Bonino}, {Brill}, {Britto}, {Bruel}, {Burnett}, {Buson}, {Cameron}, {Caputo}, {Caraveo}, {Castro}, {Chaty}, {Cheung}, {Chiaro}, {Cibrario}, {Ciprini}, {Coronado-Bl{\'a}zquez}, {Crnogorcevic}, {Cutini}, {D'Ammando}, {De Gaetano}, {Digel}, {Di Lalla}, {Dirirsa}, {Di Venere}, {Dom{\'\i}nguez}, {Fallah Ramazani}, {Fegan}, {Ferrara}, {Fiori}, {Fleischhack}, {Franckowiak}, {Fukazawa}, {Funk}, {Fusco}, {Galanti}, {Gammaldi}, {Gargano}, {Garrappa}, {Gasparrini}, {Giacchino}, {Giglietto}, {Giordano}, {Giroletti}, {Glanzman}, {Green}, {Grenier}, {Grondin}, {Guillemot}, {Guiriec}, {Gustafsson}, {Harding}, {Hays}, {Hewitt}, {Horan}, {Hou}, {J{\'o}hannesson}, {Karwin}, {Kayanoki}, {Kerr}, {Kuss}, {Landriu}, {Larsson}, {Latronico}, {Lemoine-Goumard}, {Li}, {Liodakis}, {Longo}, {Loparco}, {Lott}, {Lubrano}, {Maldera}, {Malyshev}, {Manfreda}, {Mart{\'\i}-Devesa},
  {Mazziotta}, {Mereu}, {Meyer}, {Michelson}, {Mirabal}, {Mitthumsiri}, {Mizuno}, {Moiseev}, {Monzani}, {Morselli}, {Moskalenko}, {Negro}, {Nuss}, {Omodei}, {Orienti}, {Orlando}, {Paneque}, {Pei}, {Perkins}, {Persic}, {Pesce-Rollins}, {Petrosian}, {Pillera}, {Poon}, {Porter}, {Principe}, {Rain{\`o}}, {Rando}, {Rani}, {Razzano}, {Razzaque}, {Reimer}, {Reimer}, {Reposeur}, {S{\'a}nchez-Conde}, {Saz Parkinson}, {Scotton}, {Serini}, {Sgr{\`o}}, {Siskind}, {Smith}, {Spandre}, {Spinelli}, {Sueoka}, {Suson}, {Tajima}, {Tak}, {Thayer}, {Thompson}, {Torres}, {Troja}, {Valverde}, {Wood}, \& {Zaharijas}}]{Abdollahi+22}
{Abdollahi}, S., {Acero}, F., {Baldini}, L., {et~al.} 2022, \apjs, 260, 53, \dodoi{10.3847/1538-4365/ac6751}

\bibitem[{{Ackermann} {et~al.}(2013){Ackermann}, {Ajello}, {Allafort}, {Baldini}, {Ballet}, {Barbiellini}, {Baring}, {Bastieri}, {Bechtol}, {Bellazzini}, {Blandford}, {Bloom}, {Bonamente}, {Borgland}, {Bottacini}, {Brandt}, {Bregeon}, {Brigida}, {Bruel}, {Buehler}, {Busetto}, {Buson}, {Caliandro}, {Cameron}, {Caraveo}, {Casandjian}, {Cecchi}, {{\c{C}}elik}, {Charles}, {Chaty}, {Chaves}, {Chekhtman}, {Cheung}, {Chiang}, {Chiaro}, {Cillis}, {Ciprini}, {Claus}, {Cohen-Tanugi}, {Cominsky}, {Conrad}, {Corbel}, {Cutini}, {D'Ammando}, {de Angelis}, {de Palma}, {Dermer}, {do Couto e Silva}, {Drell}, {Drlica-Wagner}, {Falletti}, {Favuzzi}, {Ferrara}, {Franckowiak}, {Fukazawa}, {Funk}, {Fusco}, {Gargano}, {Germani}, {Giglietto}, {Giommi}, {Giordano}, {Giroletti}, {Glanzman}, {Godfrey}, {Grenier}, {Grondin}, {Grove}, {Guiriec}, {Hadasch}, {Hanabata}, {Harding}, {Hayashida}, {Hayashi}, {Hays}, {Hewitt}, {Hill}, {Hughes}, {Jackson}, {Jogler}, {J{\'o}hannesson}, {Johnson}, {Kamae}, {Kataoka}, {Katsuta}, {Kn{\"o}dlseder},
  {Kuss}, {Lande}, {Larsson}, {Latronico}, {Lemoine-Goumard}, {Longo}, {Loparco}, {Lovellette}, {Lubrano}, {Madejski}, {Massaro}, {Mayer}, {Mazziotta}, {McEnery}, {Mehault}, {Michelson}, {Mignani}, {Mitthumsiri}, {Mizuno}, {Moiseev}, {Monzani}, {Morselli}, {Moskalenko}, {Murgia}, {Nakamori}, {Nemmen}, {Nuss}, {Ohno}, {Ohsugi}, {Omodei}, {Orienti}, {Orlando}, {Ormes}, {Paneque}, {Perkins}, {Pesce-Rollins}, {Piron}, {Pivato}, {Rain{\`o}}, {Rando}, {Razzano}, {Razzaque}, {Reimer}, {Reimer}, {Ritz}, {Romoli}, {S{\'a}nchez-Conde}, {Schulz}, {Sgr{\`o}}, {Simeon}, {Siskind}, {Smith}, {Spandre}, {Spinelli}, {Stecker}, {Strong}, {Suson}, {Tajima}, {Takahashi}, {Takahashi}, {Tanaka}, {Thayer}, {Thayer}, {Thompson}, {Thorsett}, {Tibaldo}, {Tibolla}, {Tinivella}, {Troja}, {Uchiyama}, {Usher}, {Vandenbroucke}, {Vasileiou}, {Vianello}, {Vitale}, {Waite}, {Werner}, {Winer}, {Wood}, {Wood}, {Yamazaki}, {Yang}, \& {Zimmer}}]{Ackermann+13}
{Ackermann}, M., {Ajello}, M., {Allafort}, A., {et~al.} 2013, Science, 339, 807, \dodoi{10.1126/science.1231160}

\bibitem[{{Aiello} {et~al.}(2024){Aiello}, {Albert}, {Alshamsi}, {Garre}, {Aly}, {Ambrosone}, {Ameli}, {Andre}, {Androutsou}, {Anguita}, {Aphecetche}, {Ardid}, {Ardid}, {Atmani}, {Aublin}, {Badaracco}, {Bailly-Salins}, {Barda{\v{c}}ov{\'a}}, {Baret}, {Bariego-Quintana}, {du Pree}, {Becherini}, {Bendahman}, {Benfenati}, {Benhassi}, {Benoit}, {Berbee}, {Bertin}, {Biagi}, {Boettcher}, {Bonanno}, {Boumaaza}, {Bouta}, {Bouwhuis}, {Bozza}, {Bozza}, {H. Br{\^a}nza{\c{s}}}, {Bretaudeau}, {Breuhaus}, {Bruijn}, {Brunner}, {Bruno}, {Buis}, {Buompane}, {Busto}, {Caiffi}, {Calvo}, {Campion}, {Capone}, {Carenini}, {Carretero}, {Cartraud}, {Castaldi}, {Cecchini}, {Celli}, {Cerisy}, {Chabab}, {Chadolias}, {Chen}, {Cherubini}, {Chiarusi}, {Circella}, {Cocimano}, {Coelho}, {Coleiro}, {Coniglione}, {Coyle}, {Creusot}, {Cuttone}, {Dallier}, {Darras}, {De Benedittis}, {De Martino}, {Decoene}, {Del Burgo}, {Del Rosso}, {Di Mauro}, {Di Palma}, {D{\'\i}az}, {Diaz}, {Diego-Tortosa}, {Distefano}, {Domi}, {Donzaud}, {Dornic},
  {D{\"o}rr}, {Drakopoulou}, {Drouhin}, {Dvornick{\'y}}, {Eberl}, {Eckerov{\'a}}, {Eddymaoui}, {van Eeden}, {Eff}, {van Eijk}, {El Bojaddaini}, {El Hedri}, {Enzenh{\"o}fer}, {Ferrara}, {Filipovi{\'c}}, {Filippini}, {Franciotti}, {Fusco}, {Gabriel}, {Gagliardini}, {Gal}, {M{\'e}ndez}, {Soto}, {Oliver}, {Gei{\ss}elbrecht}, {Ghaddari}, {Gialanella}, {Gibson}, {Giorgio}, {Goos}, {Goswami}, {Goupilliere}, {Gozzini}, {Gracia}, {Graf}, {Guidi}, {Guillon}, {Guti{\'e}rrez}, {van Haren}, {Heijboer}, {Hekalo}, {Hennig}, {Hern{\'a}ndez-Rey}, {Ibnsalih}, {Illuminati}, {de Jong}, {de Jong}, {Jung}, {Kalaczy{\'n}ski}, {Kalekin}, {Katz}, {Kistauri}, {Kopper}, {Kouchner}, {Kueviakoe}, {Kulikovskiy}, {Kvatadze}, {Labalme}, {Lahmann}, {Larosa}, {Lastoria}, {Lazo}, {Stum}, {Lehaut}, {Leonora}, {Lessing}, {Levi}, {Clark}, {Longhitano}, {Magnani}, {Majumdar}, {Malerba}, {Mamedov}, {Ma{\'n}czak}, {Manfreda}, {Manzaneda}, {Marconi}, {Margiotta}, {Marinelli}, {Markou}, {Martin}, {Mart{\'\i}nez-Mora}, {Marzaioli}, {Mastrodicasa},
  {Mastroianni}, {Miccich{\`e}}, {Miele}, {Migliozzi}, {Migneco}, {Mitsou}, {Mollo}, {Morales-Gallegos}, {Morga}, {Moussa}, {Mateo}, {Muller}, {Musone}, {Musumeci}, {Navas}, {Nayerhoda}, {Nicolau}, {Nkosi}, {Fearraigh}, {Oliviero}, {Orlando}, {Oukacha}, {Paesani}, {Gonz{\'a}lez}, {Papalashvili}, {Parisi}, {Gomez}, {P{\u{a}}un}, {P{\u{a}}v{\u{a}}la{\c{s}}}, {Mart{\'\i}nez}, {Perrin-Terrin}, {Perronnel}, {Pestel}, {Pestes}, {Piattelli}, {Poir{\`e}}, {Popa}, {Pradier}, {Prado}, {Pulvirenti}, {Qu{\'e}m{\'e}ner}, {Quiroz-Rangel}, {Rahaman}, {Randazzo}, {Randriatoamanana}, {Razzaque}, {Rea}, {Real}, {Riccobene}, {Robinson}, {Romanov}, {{\v{S}}aina}, {Greus}, {Samtleben}, {Losa}, {Sanfilippo}, {Sanguineti}, {Santonastaso}, {Santonocito}, {Sapienza}, {Schnabel}, {Schumann}, {Schutte}, {Seneca}, {Sennan}, {Setter}, {Sgura}, {Shanidze}, {Sharma}, {Shitov}, {{\v{S}}imkovic}, {Simonelli}, {Sinopoulou}, {Smirnov}, {Spisso}, {Spurio}, {Stavropoulos}, {{\v{S}}tekl}, {Taiuti}, {Tayalati}, {Thiersen}, {Tosta e Melo},
  {Tragia}, {Trocm{\'e}}, {Tsourapis}, {Tudorache}, {Tzamariudaki}, {Vacheret}, {Melchor}, {Valsecchi}, {Van Elewyck}, {Vannoye}, {Vasileiadis}, {de Sola}, {Verilhac}, {Veutro}, {Viola}, {Vivolo}, {Wilms}, {de Wolf}, {Yepes-Ramirez}, {Zarpapis}, {Zavatarelli}, {Zegarelli}, {Zito}, {Zornoza}, {Z{\'u}{\~n}iga}, \& {Zywucka}}]{Aiello+2024}
{Aiello}, S., {Albert}, A., {Alshamsi}, M., {et~al.} 2024, Astroparticle Physics, 162, 102990, \dodoi{10.1016/j.astropartphys.2024.102990}

\bibitem[{Ajello {et~al.}(2020)Ajello, Mauro, Paliya, \& Garrappa}]{Ajello+20}
Ajello, M., Mauro, M.~D., Paliya, V.~S., \& Garrappa, S. 2020, The Astrophysical Journal, 894, 88, \dodoi{10.3847/1538-4357/ab86a6}

\bibitem[{Ajello {et~al.}(2023)Ajello, Murase, \& McDaniel}]{Ajello:2023hkh}
Ajello, M., Murase, K., \& McDaniel, A. 2023, Astrophys. J. Lett., 954, L49, \dodoi{10.3847/2041-8213/acf296}

\bibitem[{{Ajello} {et~al.}(2021){Ajello}, {Baldini}, {Ballet}, {Barbiellini}, {Bastieri}, {Bellazzini}, {Berretta}, {Bissaldi}, {Blandford}, {Bloom}, {Bonino}, {Bruel}, {Buson}, {Cameron}, {Caprioli}, {Caputo}, {Cavazzuti}, {Chartas}, {Chen}, {Cheung}, {Chiaro}, {Costantin}, {Cutini}, {D'Ammando}, {de la Torre Luque}, {de Palma}, {Desai}, {Diesing}, {Di Lalla}, {Dirirsa}, {Di Venere}, {Dom{\'\i}nguez}, {Fegan}, {Franckowiak}, {Fukazawa}, {Funk}, {Fusco}, {Gargano}, {Gasparrini}, {Giglietto}, {Giordano}, {Giroletti}, {Green}, {Grenier}, {Guiriec}, {Hartmann}, {Horan}, {J{\'o}hannesson}, {Karwin}, {Kerr}, {Kova{\v{c}}evi{\'c}}, {Kuss}, {Larsson}, {Latronico}, {Lemoine-Goumard}, {Li}, {Liodakis}, {Longo}, {Loparco}, {Lovellette}, {Lubrano}, {Maldera}, {Manfreda}, {Marchesi}, {Marcotulli}, {Mart{\'\i}-Devesa}, {Mazziotta}, {Mereu}, {Michelson}, {Mizuno}, {Monzani}, {Morselli}, {Moskalenko}, {Negro}, {Omodei}, {Orienti}, {Orlando}, {Paliya}, {Paneque}, {Pei}, {Persic}, {Pesce-Rollins}, {Porter}, {Principe},
  {Racusin}, {Rain{\`o}}, {Rando}, {Rani}, {Razzano}, {Reimer}, {Reimer}, {Saz Parkinson}, {Serini}, {Sgr{\`o}}, {Siskind}, {Spandre}, {Spinelli}, {Suson}, {Tak}, {Torres}, {Troja}, {Wood}, {Zaharijas}, \& {Zrake}}]{Ajello+21}
{Ajello}, M., {Baldini}, L., {Ballet}, J., {et~al.} 2021, \apj, 921, 144, \dodoi{10.3847/1538-4357/ac1bb2}

\bibitem[{{Aramaki} {et~al.}(2020){Aramaki}, {Adrian}, {Karagiorgi}, \& {Odaka}}]{Aramaki+20}
{Aramaki}, T., {Adrian}, P. O.~H., {Karagiorgi}, G., \& {Odaka}, H. 2020, Astroparticle Physics, 114, 107, \dodoi{10.1016/j.astropartphys.2019.07.002}

\bibitem[{{Ballet} {et~al.}(2023){Ballet}, {Bruel}, {Burnett}, {Lott}, \& {The Fermi-LAT collaboration}}]{Ballet+2023}
{Ballet}, J., {Bruel}, P., {Burnett}, T.~H., {Lott}, B., \& {The Fermi-LAT collaboration}. 2023, arXiv e-prints, arXiv:2307.12546, \dodoi{10.48550/arXiv.2307.12546}

\bibitem[{{Beckmann} {et~al.}(2009){Beckmann}, {Soldi}, {Ricci}, {Alfonso-Garz{\'o}n}, {Courvoisier}, {Domingo}, {Gehrels}, {Lubi{\'n}ski}, {Mas-Hesse}, \& {Zdziarski}}]{Beckmann+2009}
{Beckmann}, V., {Soldi}, S., {Ricci}, C., {et~al.} 2009, \aap, 505, 417, \dodoi{10.1051/0004-6361/200912111}

\bibitem[{{Bell}(1978{\natexlab{a}})}]{Bell78a}
{Bell}, A.~R. 1978{\natexlab{a}}, \mnras, 182, 147, \dodoi{10.1093/mnras/182.2.147}

\bibitem[{{Bell}(1978{\natexlab{b}})}]{Bell78b}
---. 1978{\natexlab{b}}, \mnras, 182, 443, \dodoi{10.1093/mnras/182.3.443}

\bibitem[{{Bird} {et~al.}(2007){Bird}, {Malizia}, {Bazzano}, {Barlow}, {Bassani}, {Hill}, {B{\'e}langer}, {Capitanio}, {Clark}, {Dean}, {Fiocchi}, {G{\"o}tz}, {Lebrun}, {Molina}, {Produit}, {Renaud}, {Sguera}, {Stephen}, {Terrier}, {Ubertini}, {Walter}, {Winkler}, \& {Zurita}}]{Bird+2007}
{Bird}, A.~J., {Malizia}, A., {Bazzano}, A., {et~al.} 2007, \apjs, 170, 175, \dodoi{10.1086/513148}

\bibitem[{{Blandford} \& {Ostriker}(1978)}]{Blandford&Ostriker78}
{Blandford}, R.~D., \& {Ostriker}, J.~P. 1978, \apjl, 221, L29, \dodoi{10.1086/182658}

\bibitem[{Blumenthal(1970)}]{Blumenthal70}
Blumenthal, G.~R. 1970, Phys. Rev. D, 1, 1596, \dodoi{10.1103/PhysRevD.1.1596}

\bibitem[{{Blumenthal} \& {Gould}(1970)}]{Blumenthal&Gould70}
{Blumenthal}, G.~R., \& {Gould}, R.~J. 1970, Reviews of Modern Physics, 42, 237, \dodoi{10.1103/RevModPhys.42.237}

\bibitem[{{Blustin} {et~al.}(2005){Blustin}, {Page}, {Fuerst}, {Branduardi-Raymont}, \& {Ashton}}]{Blustin+2005}
{Blustin}, A.~J., {Page}, M.~J., {Fuerst}, S.~V., {Branduardi-Raymont}, G., \& {Ashton}, C.~E. 2005, \aap, 431, 111, \dodoi{10.1051/0004-6361:20041775}

\bibitem[{{Boissay} {et~al.}(2016){Boissay}, {Ricci}, \& {Paltani}}]{Boissay+2016A&A...588A..70B}
{Boissay}, R., {Ricci}, C., \& {Paltani}, S. 2016, \aap, 588, A70, \dodoi{10.1051/0004-6361/201526982}

\bibitem[{{Bolatto} {et~al.}(2013){Bolatto}, {Wolfire}, \& {Leroy}}]{2013ARA&A..51..207B}
{Bolatto}, A.~D., {Wolfire}, M., \& {Leroy}, A.~K. 2013, \araa, 51, 207, \dodoi{10.1146/annurev-astro-082812-140944}

\bibitem[{Buson {et~al.}(2023)Buson, Tramacere, Oswald, Barbano, de~Clairfontaine, Pfeiffer, Azzollini, Baghmanyan, \& Ajello}]{Buson:2023irp}
Buson, S., Tramacere, A., Oswald, L., {et~al.} 2023.
\newblock \doarXiv{2305.11263}

\bibitem[{{Cherenkov Telescope Array Consortium} {et~al.}(2019){Cherenkov Telescope Array Consortium}, {Acharya}, {Agudo}, {Al Samarai}, {Alfaro}, {Alfaro}, {Alispach}, {Alves Batista}, {Amans}, {Amato}, {Ambrosi}, {Antolini}, {Antonelli}, {Aramo}, {Araya}, {Armstrong}, {Arqueros}, {Arrabito}, {Asano}, {Ashley}, {Backes}, {Balazs}, {Balbo}, {Ballester}, {Ballet}, {Bamba}, {Barkov}, {Barres de Almeida}, {Barrio}, {Bastieri}, {Becherini}, {Belfiore}, {Benbow}, {Berge}, {Bernardini}, {Bernardini}, {Bernardos}, {Bernl{\"o}hr}, {Bertucci}, {Biasuzzi}, {Bigongiari}, {Biland}, {Bissaldi}, {Biteau}, {Blanch}, {Blazek}, {Boisson}, {Bolmont}, {Bonanno}, {Bonardi}, {Bonavolont{\`a}}, {Bonnoli}, {Bosnjak}, {B{\"o}ttcher}, {Braiding}, {Bregeon}, {Brill}, {Brown}, {Brun}, {Brunetti}, {Buanes}, {Buckley}, {Bugaev}, {B{\"u}hler}, {Bulgarelli}, {Bulik}, {Burton}, {Burtovoi}, {Busetto}, {Canestrari}, {Capalbi}, {Capitanio}, {Caproni}, {Caraveo}, {C{\'a}rdenas}, {Carlile}, {Carosi}, {Carqu{\'\i}n}, {Carr}, {Casanova},
  {Cascone}, {Catalani}, {Catalano}, {Cauz}, {Cerruti}, {Chadwick}, {Chaty}, {Chaves}, {Chen}, {Chen}, {Chernyakova}, {Chikawa}, {Christov}, {Chudoba}, {Cie{\'s}lar}, {Coco}, {Colafrancesco}, {Colin}, {Conforti}, {Connaughton}, {Conrad}, {Contreras}, {Cortina}, {Costa}, {Costantini}, {Cotter}, {Covino}, {Crocker}, {Cuadra}, {Cuevas}, {Cumani}, {D'A{\`\i}}, {D'Ammando}, {D'Avanzo}, {D'Urso}, {Daniel}, {Davids}, {Dawson}, {Dazzi}, {De Angelis}, {de C{\'a}ssia dos Anjos}, {De Cesare}, {De Franco}, {de Gouveia Dal Pino}, {de la Calle}, {de los Reyes Lopez}, {De Lotto}, {De Luca}, {De Lucia}, {de Naurois}, {de O{\~n}a Wilhelmi}, {De Palma}, {De Persio}, {de Souza}, {Deil}, {Del Santo}, {Delgado}, {della Volpe}, {Di Girolamo}, {Di Pierro}, {Di Venere}, {D{\'\i}az}, {Dib}, {Diebold}, {Djannati-Ata{\"\i}}, {Dom{\'\i}nguez}, {Dominis Prester}, {Dorner}, {Doro}, {Drass}, {Dravins}, {Dubus}, {Dwarkadas}, {Ebr}, {Eckner}, {Egberts}, {Einecke}, {Ekoume}, {Els{\"a}sser}, {Ernenwein}, {Espinoza}, {Evoli}, {Fairbairn},
  {Falceta-Goncalves}, {Falcone}, {Farnier}, {Fasola}, {Fedorova}, {Fegan}, {Fernandez-Alonso}, {Fern{\'a}ndez-Barral}, {Ferrand}, {Fesquet}, {Filipovic}, {Fioretti}, {Fontaine}, {Fornasa}, {Fortson}, {Freixas Coromina}, {Fruck}, {Fujita}, {Fukazawa}, {Funk}, {F{\"u}{\ss}ling}, {Gabici}, {Gadola}, {Gallant}, {Garcia}, {Garcia L{\'o}pez}, {Garczarczyk}, {Gaskins}, {Gasparetto}, {Gaug}, {Gerard}, {Giavitto}, {Giglietto}, {Giommi}, {Giordano}, {Giro}, {Giroletti}, {Giuliani}, {Glicenstein}, {Gnatyk}, {Godinovic}, {Goldoni}, {G{\'o}mez-Vargas}, {Gonz{\'a}lez}, {Gonz{\'a}lez}, {G{\"o}tz}, {Graham}, {Grandi}, {Granot}, {Green}, {Greenshaw}, {Griffiths}, {Gunji}, {Hadasch}, {Hara}, {Hardcastle}, {Hassan}, {Hayashi}, {Hayashida}, {Heller}, {Helo}, {Hermann}, {Hinton}, {Hnatyk}, {Hofmann}, {Holder}, {Horan}, {H{\"o}randel}, {Horns}, {Horvath}, {Hovatta}, {Hrabovsky}, {Hrupec}, {Humensky}, {H{\"u}tten}, {Iarlori}, {Inada}, {Inome}, {Inoue}, {Inoue}, {Inoue}, {Iocco}, {Ioka}, {Iori}, {Ishio}, {Iwamura}, {Jamrozy},
  {Janecek}, {Jankowsky}, {Jean}, {Jung-Richardt}, {Jurysek}, {Kaaret}, {Karkar}, {Katagiri}, {Katz}, {Kawanaka}, {Kazanas}, {Kh{\'e}lifi}, {Kieda}, {Kimeswenger}, {Kimura}, {Kisaka}, {Knapp}, {Kn{\"o}dlseder}, {Koch}, {Kohri}, {Komin}, {Kosack}, {Kraus}, {Krause}, {Krau{\ss}}, {Kubo}, {Kukec Mezek}, {Kuroda}, {Kushida}, {La Palombara}, {Lamanna}, {Lang}, {Lapington}, {Le Blanc}, {Leach}, {Lees}, {Lefaucheur}, {Leigui de Oliveira}, {Lenain}, {Lico}, {Limon}, {Lindfors}, {Lohse}, {Lombardi}, {Longo}, {L{\'o}pez}, {L{\'o}pez-Coto}, {Lu}, {Lucarelli}, {Luque-Escamilla}, {Lyard}, {Maccarone}, {Maier}, {Majumdar}, {Malaguti}, {Mandat}, {Maneva}, {Manganaro}, {Mangano}, {Marcowith}, {Mar{\'\i}n}, {Markoff}, {Mart{\'\i}}, {Martin}, {Mart{\'\i}nez}, {Mart{\'\i}nez}, {Masetti}, {Masuda}, {Maurin}, {Maxted}, {Mazin}, {Medina}, {Melandri}, {Mereghetti}, {Meyer}, {Minaya}, {Mirabal}, {Mirzoyan}, {Mitchell}, {Mizuno}, {Moderski}, {Mohammed}, {Mohrmann}, {Montaruli}, {Moralejo}, {Morcuende-Parrilla}, {Mori}, {Morlino},
  {Morris}, {Morselli}, {Moulin}, {Mukherjee}, {Mundell}, {Murach}, {Muraishi}, {Murase}, {Nagai}, {Nagataki}, {Nagayoshi}, {Naito}, {Nakamori}, {Nakamura}, {Niemiec}, {Nieto}, {Niko{\l}ajuk}, {Nishijima}, {Noda}, {Nosek}, {Novosyadlyj}, {Nozaki}, {O'Brien}, {Oakes}, {Ohira}, {Ohishi}, {Ohm}, {Okazaki}, {Okumura}, {Ong}, {Orienti}, {Orito}, {Osborne}, {Ostrowski}, {Otte}, {Oya}, {Padovani}, {Paizis}, {Palatiello}, {Palatka}, {Paoletti}, {Paredes}, {Pareschi}, {Parsons}, {Pe'er}, {Pech}, {Pedaletti}, {Perri}, {Persic}, {Petrashyk}, {Petrucci}, {Petruk}, {Peyaud}, {Pfeifer}, {Piano}, {Pisarski}, {Pita}, {Pohl}, {Polo}, {Pozo}, {Prandini}, {Prast}, {Principe}, {Prokhorov}, {Prokoph}, {Prouza}, {P{\"u}hlhofer}, {Punch}, {P{\"u}rckhauer}, {Queiroz}, {Quirrenbach}, {Rain{\`o}}, {Razzaque}, {Reimer}, {Reimer}, {Reisenegger}, {Renaud}, {Rezaeian}, {Rhode}, {Ribeiro}, {Rib{\'o}}, {Richtler}, {Rico}, {Rieger}, {Riquelme}, {Rivoire}, {Rizi}, {Rodriguez}, {Rodriguez Fernandez}, {Rodr{\'\i}guez V{\'a}zquez}, {Rojas},
  {Romano}, {Romeo}, {Rosado}, {Rovero}, {Rowell}, {Rudak}, {Rugliancich}, {Rulten}, {Sadeh}, {Safi-Harb}, {Saito}, {Sakaki}, {Sakurai}, {Salina}, {S{\'a}nchez-Conde}, {Sandaker}, {Sandoval}, {Sangiorgi}, {Sanguillon}, {Sano}, {Santander}, {Sarkar}, {Satalecka}, {Saturni}, {Schioppa}, {Schlenstedt}, {Schneider}, {Schoorlemmer}, {Schovanek}, {Schulz}, {Schussler}, {Schwanke}, {Sciacca}, {Scuderi}, {Seitenzahl}, {Semikoz}, {Sergijenko}, {Servillat}, {Shalchi}, {Shellard}, {Sidoli}, {Siejkowski}, {Sillanp{\"a}{\"a}}, {Sironi}, {Sitarek}, {Sliusar}, {Slowikowska}, {Sol}, {Stamerra}, {Stani{\v{c}}}, {Starling}, {Stawarz}, {Stefanik}, {Stephan}, {Stolarczyk}, {Stratta}, {Straumann}, {Suomijarvi}, {Supanitsky}, {Tagliaferri}, {Tajima}, {Tavani}, {Tavecchio}, {Tavernet}, {Tayabaly}, {Tejedor}, {Temnikov}, {Terada}, {Terrier}, {Terzic}, {Teshima}, {Testa}, {Thoudam}, {Tian}, {Tibaldo}, {Tluczykont}, {Todero Peixoto}, {Tokanai}, {Tomastik}, {Tonev}, {Tornikoski}, {Torres}, {Torresi}, {Tosti}, {Tothill}, {Tovmassian},
  {Travnicek}, {Trichard}, {Trifoglio}, {Troyano Pujadas}, {Tsujimoto}, {Umana}, {Vagelli}, {Vagnetti}, {Valentino}, {Vallania}, {Valore}, {van Eldik}, {Vandenbroucke}, {Varner}, {Vasileiadis}, {Vassiliev}, {V{\'a}zquez Acosta}, {Vecchi}, {Vega}, {Vercellone}, {Veres}, {Vergani}, {Verzi}, {Vettolani}, {Viana}, {Vigorito}, {Villanueva}, {Voelk}, {Vollhardt}, {Vorobiov}, {Vrastil}, {Vuillaume}, {Wagner}, {Wagner}, {Walter}, {Ward}, {Warren}, {Watson}, {Werner}, {White}, {White}, {Wierzcholska}, {Wilcox}, {Will}, {Williams}, {Wischnewski}, {Wood}, {Yamamoto}, {Yamazaki}, {Yanagita}, {Yang}, {Yoshida}, {Yoshiike}, {Yoshikoshi}, {Zacharias}, {Zaharijas}, {Zampieri}, {Zandanel}, {Zanin}, {Zavrtanik}, {Zavrtanik}, {Zdziarski}, {Zech}, {Zechlin}, {Zhdanov}, {Ziegler}, \& {Zorn}}]{CTA+2019}
{Cherenkov Telescope Array Consortium}, {Acharya}, B.~S., {Agudo}, I., {et~al.} 2019, {Science with the Cherenkov Telescope Array}, \dodoi{10.1142/10986}

\bibitem[{{Condon} {et~al.}(1998){Condon}, {Cotton}, {Greisen}, {Yin}, {Perley}, {Taylor}, \& {Broderick}}]{Condon+1998}
{Condon}, J.~J., {Cotton}, W.~D., {Greisen}, E.~W., {et~al.} 1998, \aj, 115, 1693, \dodoi{10.1086/300337}

\bibitem[{{Cusumano} {et~al.}(2010){Cusumano}, {La Parola}, {Segreto}, {Ferrigno}, {Maselli}, {Sbarufatti}, {Romano}, {Chincarini}, {Giommi}, {Masetti}, {Moretti}, {Parisi}, \& {Tagliaferri}}]{Cusumano+2010}
{Cusumano}, G., {La Parola}, V., {Segreto}, A., {et~al.} 2010, \aap, 524, A64, \dodoi{10.1051/0004-6361/201015249}

\bibitem[{{Douglas} {et~al.}(1996){Douglas}, {Bash}, {Bozyan}, {Torrence}, \& {Wolfe}}]{Douglas+1996}
{Douglas}, J.~N., {Bash}, F.~N., {Bozyan}, F.~A., {Torrence}, G.~W., \& {Wolfe}, C. 1996, \aj, 111, 1945, \dodoi{10.1086/117932}

\bibitem[{{Drury}(1983)}]{Drury+83}
{Drury}, L.~O. 1983, Reports on Progress in Physics, 46, 973, \dodoi{10.1088/0034-4885/46/8/002}

\bibitem[{{Ehlert} {et~al.}(2024){Ehlert}, {Oikonomou}, \& {Peretti}}]{2024arXiv241105667E}
{Ehlert}, D., {Oikonomou}, F., \& {Peretti}, E. 2024, arXiv e-prints, arXiv:2411.05667, \dodoi{10.48550/arXiv.2411.05667}

\bibitem[{Eichmann {et~al.}(2022)Eichmann, Oikonomou, Salvatore, Dettmar, \& Becker~Tjus}]{Eichmann:2022lxh}
Eichmann, B., Oikonomou, F., Salvatore, S., Dettmar, R.-J., \& Becker~Tjus, J. 2022, Astrophys. J., 939, 43, \dodoi{10.3847/1538-4357/ac9588}

\bibitem[{{Fang} {et~al.}(2023){Fang}, {Lopez Rodriguez}, {Halzen}, \& {Gallagher}}]{Fang+2024}
{Fang}, K., {Lopez Rodriguez}, E., {Halzen}, F., \& {Gallagher}, J.~S. 2023, \apj, 956, 8, \dodoi{10.3847/1538-4357/acee70}

\bibitem[{{Faucher-Gigu{\`e}re} \& {Quataert}(2012)}]{Andre+12}
{Faucher-Gigu{\`e}re}, C.-A., \& {Quataert}, E. 2012, \mnras, 425, 605, \dodoi{10.1111/j.1365-2966.2012.21512.x}

\bibitem[{{Fukazawa} {et~al.}(2022){Fukazawa}, {Matake}, {Kayanoki}, {Inoue}, \& {Finke}}]{Fukazawa+22}
{Fukazawa}, Y., {Matake}, H., {Kayanoki}, T., {Inoue}, Y., \& {Finke}, J. 2022, \apj, 931, 138, \dodoi{10.3847/1538-4357/ac6acb}

\bibitem[{{Ginzburg} \& {Syrovatski{\u{i}}}(1966)}]{Ginzburg+66}
{Ginzburg}, V.~L., \& {Syrovatski{\u{i}}}, S.~I. 1966, Soviet Physics Uspekhi, 9, 223, \dodoi{10.1070/PU1966v009n02ABEH002871}

\bibitem[{Hooper \& Plant(2023)}]{Hooper:2023ssc}
Hooper, D., \& Plant, K. 2023, Phys. Rev. Lett., 131, 231001, \dodoi{10.1103/PhysRevLett.131.231001}

\bibitem[{Huang {et~al.}(2024)Huang, Wang, \& Ma}]{Huang:2024yua}
Huang, Y.-H., Wang, K., \& Ma, Z.-P. 2024.
\newblock \doarXiv{2406.14001}

\bibitem[{{Huentemeyer} {et~al.}(2019){Huentemeyer}, {BenZvi}, {Dingus}, {Fleischhack}, {Schoorlemmer}, \& {Weisgarber}}]{Huentemeyer+2019}
{Huentemeyer}, P., {BenZvi}, S., {Dingus}, B., {et~al.} 2019, in Bulletin of the American Astronomical Society, Vol.~51, 109, \dodoi{10.48550/arXiv.1907.07737}

\bibitem[{{IceCube Collaboration} {et~al.}(2022){IceCube Collaboration}, {Abbasi}, {Ackermann}, {Adams}, {Aguilar}, {Ahlers}, {Ahrens}, {Alameddine}, {Alispach}, {Alves}, {Amin}, {Andeen}, {Anderson}, {Anton}, {Arg{\"u}elles}, {Ashida}, {Axani}, {Bai}, {Balagopal}, {Barbano}, {Barwick}, {Bastian}, {Basu}, {Baur}, {Bay}, {Beatty}, {Becker}, {Becker Tjus}, {Bellenghi}, {Benzvi}, {Berley}, {Bernardini}, {Besson}, {Binder}, {Bindig}, {Blaufuss}, {Blot}, {Boddenberg}, {Bontempo}, {Borowka}, {B{\"o}ser}, {Botner}, {B{\"o}ttcher}, {Bourbeau}, {Bradascio}, {Braun}, {Brinson}, {Bron}, {Brostean-Kaiser}, {Browne}, {Burgman}, {Burley}, {Busse}, {Campana}, {Carnie-Bronca}, {Chen}, {Chen}, {Chirkin}, {Choi}, {Clark}, {Clark}, {Classen}, {Coleman}, {Collin}, {Conrad}, {Coppin}, {Correa}, {Cowen}, {Cross}, {Dappen}, {Dave}, {de Clercq}, {Delaunay}, {Delgado L{\'o}pez}, {Dembinski}, {Deoskar}, {Desai}, {Desiati}, {de Vries}, {de Wasseige}, {de With}, {Deyoung}, {Diaz}, {D{\'\i}az-V{\'e}lez}, {Dittmer}, {Dujmovic}, {Dunkman},
  {Duvernois}, {Dvorak}, {Ehrhardt}, {Eller}, {Engel}, {Erpenbeck}, {Evans}, {Evenson}, {Fan}, {Fazely}, {Fedynitch}, {Feigl}, {Fiedlschuster}, {Fienberg}, {Filimonov}, {Finley}, {Fischer}, {Fox}, {Franckowiak}, {Friedman}, {Fritz}, {F{\"u}rst}, {Gaisser}, {Gallagher}, {Ganster}, {Garcia}, {Garrappa}, {Gerhardt}, {Ghadimi}, {Glaser}, {Glauch}, {Gl{\"u}senkamp}, {Goldschmidt}, {Gonzalez}, {Goswami}, {Grant}, {Gr{\'e}goire}, {Griswold}, {G{\"u}nther}, {Gutjahr}, {Haack}, {Hallgren}, {Halliday}, {Halve}, {Halzen}, {Hanson}, {Hardin}, {Harnisch}, {Haungs}, {Hebecker}, {Helbing}, {Henningsen}, {Hettinger}, {Hickford}, {Hignight}, {Hill}, {Hill}, {Hoffman}, {Hoffmann}, {Hokanson-Fasig}, {Hoshina}, {Huang}, {Huber}, {Huber}, {Hultqvist}, {H{\"u}nnefeld}, {Hussain}, {Hymon}, {in}, {Iovine}, {Ishihara}, {Jansson}, {Japaridze}, {Jeong}, {Jin}, {Jones}, {Kang}, {Kang}, {Kang}, {Kappes}, {Kappesser}, {Kardum}, {Karg}, {Karl}, {Karle}, {Katz}, {Kauer}, {Kellermann}, {Kelley}, {Kheirandish}, {Kin}, {Kintscher}, {Kiryluk},
  {Klein}, {Koirala}, {Kolanoski}, {Kontrimas}, {K{\"o}pke}, {Kopper}, {Kopper}, {Koskinen}, {Koundal}, {Kovacevich}, {Kowalski}, {Kozynets}, {Kun}, {Kurahashi}, {Lad}, {Lagunas Gualda}, {Lanfranchi}, {Larson}, {Lauber}, {Lazar}, {Lee}, {Leonard}, {Leszczy{\'n}ska}, {Li}, {Lincetto}, {Liu}, {Liubarska}, {Lohfink}, {Lozano Mariscal}, {Lu}, {Lucarelli}, {Ludwig}, {Luszczak}, {Lyu}, {Ma}, {Madsen}, {Mahn}, {Makino}, {Mancina}, {Mari{\c{s}}}, {Martinez-Soler}, {Maruyama}, {Mase}, {McElroy}, {McNally}, {Mead}, {Meagher}, {Mechbal}, {Medina}, {Meier}, {Meighen-Berger}, {Micallef}, {Mockler}, {Montaruli}, {Moore}, {Morse}, {Moulai}, {Naab}, {Nagai}, {Nahnhauer}, {Naumann}, {Necker}, {Nguyen}, {Niederhausen}, {Nisa}, {Nowicki}, {Nygren}, {Obertack}, {Pollmann}, {Oehler}, {Oeyen}, {Olivas}, {O'Sullivan}, {Pandya}, {Pankova}, {Park}, {Parker}, {Paudel}, {Paul}, {P{\'e}rez de Los Heros}, {Peters}, {Peterson}, {Philippen}, {Pieper}, {Pittermann}, {Pizzuto}, {Plum}, {Popovych}, {Porcelli}, {Prado Rodriguez}, {Price},
  {Pries}, {Przybylski}, {Rack-Helleis}, {Raissi}, {Rameez}, {Rawlins}, {Rea}, {Rehman}, {Reichherzer}, {Reimann}, {Renzi}, {Resconi}, {Reusch}, {Rhode}, {Richman}, {Riedel}, {Roberts}, {Robertson}, {Roellinghoff}, {Rongen}, {Rott}, {Ruhe}, {Ryckbosch}, {Rysewyk Cantu}, {Safa}, {Saffer}, {Sanchez Herrera}, {Sandrock}, {Sandroos}, {Santander}, {Sarkar}, {Sarkar}, {Satalecka}, {Schaufel}, {Schieler}, {Schindler}, {Schmidt}, {Schneider}, {Schneider}, {Schr{\"o}der}, {Schumacher}, {Schwefer}, {Sclafani}, {Seckel}, {Seunarine}, {Sharma}, {Shefali}, {Silva}, {Skrzypek}, {Smithers}, {Snihur}, {Soedingrekso}, {Soldin}, {Spannfellner}, {Spiczak}, {Spiering}, {Stachurska}, {Stamatikos}, {Stanev}, {Stein}, {Stettner}, {Steuer}, {Stezelberger}, {Stokstad}, {St{\"u}rwald}, {Stuttard}, {Sullivan}, {Taboada}, {Ter-Antonyan}, {Tilav}, {Tischbein}, {Tollefson}, {T{\"o}nnis}, {Toscano}, {Tosi}, {Trettin}, {Tselengidou}, {Tung}, {Turcati}, {Turcotte}, {Turley}, {Twagirayezu}, {Ty}, {Unland Elorrieta}, {Valtonen-Mattila},
  {Vandenbroucke}, {van Eijndhoven}, {Vannerom}, {van Santen}, {Verpoest}, {Walck}, {Watson}, {Weaver}, {Weigel}, {Weindl}, {Weiss}, {Weldert}, {Wendt}, {Werthebach}, {Weyrauch}, {Whitehorn}, {Wiebusch}, {Williams}, {Wolf}, {Woschnagg}, {Wrede}, {Wulff}, {Xu}, {Yanez}, {Yoshida}, {Yu}, {Yuan}, {Zhangan}, \& {Zhelnin}}]{IceCube+22}
{IceCube Collaboration}, {Abbasi}, R., {Ackermann}, M., {et~al.} 2022, Science, 378, 538, \dodoi{10.1126/science.abg3395}

\bibitem[{{Inoue} {et~al.}(2022){Inoue}, {Cerruti}, {Murase}, \& {Liu}}]{Inoue+22}
{Inoue}, S., {Cerruti}, M., {Murase}, K., \& {Liu}, R.-Y. 2022, arXiv e-prints, arXiv:2207.02097, \dodoi{10.48550/arXiv.2207.02097}

\bibitem[{Inoue(2011)}]{Inoue:2011bm}
Inoue, Y. 2011, Astrophys. J., 733, 66, \dodoi{10.1088/0004-637X/733/1/66}

\bibitem[{{Inoue} {et~al.}(2013){Inoue}, {Inoue}, {Kobayashi}, {Makiya}, {Niino}, \& {Totani}}]{Inoue+13a}
{Inoue}, Y., {Inoue}, S., {Kobayashi}, M. A.~R., {et~al.} 2013, \apj, 768, 197, \dodoi{10.1088/0004-637X/768/2/197}

\bibitem[{Inoue \& Khangulyan(2023)}]{Inoue:2023bmy}
Inoue, Y., \& Khangulyan, D. 2023, Publ. Astron. Soc. Jap., 75, L33, \dodoi{10.1093/pasj/psad072}

\bibitem[{{Inoue} {et~al.}(2020){Inoue}, {Khangulyan}, \& {Doi}}]{Inoue2020ApJ...891L..33I}
{Inoue}, Y., {Khangulyan}, D., \& {Doi}, A. 2020, \apjl, 891, L33, \dodoi{10.3847/2041-8213/ab7661}

\bibitem[{{Inoue} {et~al.}(2019){Inoue}, {Khangulyan}, {Inoue}, \& {Doi}}]{Inoue:2019fil}
{Inoue}, Y., {Khangulyan}, D., {Inoue}, S., \& {Doi}, A. 2019, \apj, 880, 40, \dodoi{10.3847/1538-4357/ab2715}

\bibitem[{{Inoue} {et~al.}(2024){Inoue}, {Takasao}, \& {Khangulyan}}]{Inoue:2024nap}
{Inoue}, Y., {Takasao}, S., \& {Khangulyan}, D. 2024, \pasj, 76, 996, \dodoi{10.1093/pasj/psae065}

\bibitem[{{Kelner} \& {Aharonian}(2008)}]{Kelner+08}
{Kelner}, S.~R., \& {Aharonian}, F.~A. 2008, \prd, 78, 034013, \dodoi{10.1103/PhysRevD.78.034013}

\bibitem[{{Kelner} {et~al.}(2006){Kelner}, {Aharonian}, \& {Bugayov}}]{Kelner+06}
{Kelner}, S.~R., {Aharonian}, F.~A., \& {Bugayov}, V.~V. 2006, \prd, 74, 034018, \dodoi{10.1103/PhysRevD.74.034018}

\bibitem[{{Khangulyan} {et~al.}(2014){Khangulyan}, {Aharonian}, \& {Kelner}}]{Khangulyan+14}
{Khangulyan}, D., {Aharonian}, F.~A., \& {Kelner}, S.~R. 2014, \apj, 783, 100, \dodoi{10.1088/0004-637X/783/2/100}

\bibitem[{{Koo} \& {McKee}(1992{\natexlab{a}})}]{Koo&McKee1992ApJ...388...93K}
{Koo}, B.-C., \& {McKee}, C.~F. 1992{\natexlab{a}}, \apj, 388, 93, \dodoi{10.1086/171132}

\bibitem[{{Koo} \& {McKee}(1992{\natexlab{b}})}]{Koo&McKee1992b}
---. 1992{\natexlab{b}}, \apj, 388, 103, \dodoi{10.1086/171133}

\bibitem[{{Krivonos} {et~al.}(2015){Krivonos}, {Tsygankov}, {Lutovinov}, {Revnivtsev}, {Churazov}, \& {Sunyaev}}]{Krivonos+2015}
{Krivonos}, R., {Tsygankov}, S., {Lutovinov}, A., {et~al.} 2015, \mnras, 448, 3766, \dodoi{10.1093/mnras/stv150}

\bibitem[{{Lacy} {et~al.}(1994){Lacy}, {Knacke}, {Geballe}, \& {Tokunaga}}]{Lacy+1994}
{Lacy}, J.~H., {Knacke}, R., {Geballe}, T.~R., \& {Tokunaga}, A.~T. 1994, \apjl, 428, L69, \dodoi{10.1086/187395}

\bibitem[{{Lamastra} {et~al.}(2016){Lamastra}, {Fiore}, {Guetta}, {Antonelli}, {Colafrancesco}, {Menci}, {Puccetti}, {Stamerra}, \& {Zappacosta}}]{Lamastra+16}
{Lamastra}, A., {Fiore}, F., {Guetta}, D., {et~al.} 2016, \aap, 596, A68, \dodoi{10.1051/0004-6361/201628667}

\bibitem[{{Lenain} {et~al.}(2010){Lenain}, {Ricci}, {T{\"u}rler}, {Dorner}, \& {Walter}}]{Lenain+2010}
{Lenain}, J.~P., {Ricci}, C., {T{\"u}rler}, M., {Dorner}, D., \& {Walter}, R. 2010, \aap, 524, A72, \dodoi{10.1051/0004-6361/201015644}

\bibitem[{{Liu} {et~al.}(2018){Liu}, {Murase}, {Inoue}, {Ge}, \& {Wang}}]{Liu+18}
{Liu}, R.-Y., {Murase}, K., {Inoue}, S., {Ge}, C., \& {Wang}, X.-Y. 2018, \apj, 858, 9, \dodoi{10.3847/1538-4357/aaba74}

\bibitem[{{Malecki}(2024)}]{Malecki2024}
{Malecki}, P. 2024, Universe, 10, 53, \dodoi{10.3390/universe10020053}

\bibitem[{{Malizia} {et~al.}(2009){Malizia}, {Stephen}, {Bassani}, {Bird}, {Panessa}, \& {Ubertini}}]{Malizia+2009}
{Malizia}, A., {Stephen}, J.~B., {Bassani}, L., {et~al.} 2009, \mnras, 399, 944, \dodoi{10.1111/j.1365-2966.2009.15330.x}

\bibitem[{{Marti} {et~al.}(1998){Marti}, {Mirabel}, {Chaty}, \& {Rodriguez}}]{Marti+98}
{Marti}, J., {Mirabel}, I.~F., {Chaty}, S., \& {Rodriguez}, L.~F. 1998, \aap, 330, 72, \dodoi{10.48550/arXiv.astro-ph/9710136}

\bibitem[{{Mel{\'e}ndez} {et~al.}(2014){Mel{\'e}ndez}, {Mushotzky}, {Shimizu}, {Barger}, \& {Cowie}}]{Melendez+2014}
{Mel{\'e}ndez}, M., {Mushotzky}, R.~F., {Shimizu}, T.~T., {Barger}, A.~J., \& {Cowie}, L.~L. 2014, \apj, 794, 152, \dodoi{10.1088/0004-637X/794/2/152}

\bibitem[{{Michiyama} {et~al.}(2024){Michiyama}, {Inoue}, {Doi}, {Yamada}, {Fukazawa}, {Kubo}, \& {Barnier}}]{Michiyama+24}
{Michiyama}, T., {Inoue}, Y., {Doi}, A., {et~al.} 2024, \apj, 965, 68, \dodoi{10.3847/1538-4357/ad2fae}

\bibitem[{{Mizumoto} {et~al.}(2019){Mizumoto}, {Izumi}, \& {Kohno}}]{Mizumoto+19}
{Mizumoto}, M., {Izumi}, T., \& {Kohno}, K. 2019, \apj, 871, 156, \dodoi{10.3847/1538-4357/aaf814}

\bibitem[{{Molinari} {et~al.}(2011){Molinari}, {Bally}, {Noriega-Crespo}, {Compi{\`e}gne}, {Bernard}, {Paradis}, {Martin}, {Testi}, {Barlow}, {Moore}, {Plume}, {Swinyard}, {Zavagno}, {Calzoletti}, {Di Giorgio}, {Elia}, {Faustini}, {Natoli}, {Pestalozzi}, {Pezzuto}, {Piacentini}, {Polenta}, {Polychroni}, {Schisano}, {Traficante}, {Veneziani}, {Battersby}, {Burton}, {Carey}, {Fukui}, {Li}, {Lord}, {Morgan}, {Motte}, {Schuller}, {Stringfellow}, {Tan}, {Thompson}, {Ward-Thompson}, {White}, \& {Umana}}]{Molinari2011ApJ...735L..33M}
{Molinari}, S., {Bally}, J., {Noriega-Crespo}, A., {et~al.} 2011, \apjl, 735, L33, \dodoi{10.1088/2041-8205/735/2/L33}

\bibitem[{{Murase} {et~al.}(2024){Murase}, {Karwin}, {Kimura}, {Ajello}, \& {Buson}}]{Murase2024ApJ...961L..34M}
{Murase}, K., {Karwin}, C.~M., {Kimura}, S.~S., {Ajello}, M., \& {Buson}, S. 2024, \apjl, 961, L34, \dodoi{10.3847/2041-8213/ad19c5}

\bibitem[{Murase {et~al.}(2020)Murase, Kimura, \& Meszaros}]{Murase:2019vdl}
Murase, K., Kimura, S.~S., \& Meszaros, P. 2020, Phys. Rev. Lett., 125, 011101, \dodoi{10.1103/PhysRevLett.125.011101}

\bibitem[{{NASA/IPAC Extragalactic Database (NED)}(2019)}]{https://doi.org/10.26132/ned1}
{NASA/IPAC Extragalactic Database (NED)}. 2019, NASA/IPAC Extragalactic Database (NED),  IPAC, \dodoi{10.26132/NED1}

\bibitem[{{Netzer}(2015)}]{2015ARA&A..53..365N}
{Netzer}, H. 2015, \araa, 53, 365, \dodoi{10.1146/annurev-astro-082214-122302}

\bibitem[{{Nims} {et~al.}(2015){Nims}, {Quataert}, \& {Faucher-Gigu{\`e}re}}]{Nims+15}
{Nims}, J., {Quataert}, E., \& {Faucher-Gigu{\`e}re}, C.-A. 2015, \mnras, 447, 3612, \dodoi{10.1093/mnras/stu2648}

\bibitem[{{Nomura} {et~al.}(2016){Nomura}, {Ohsuga}, {Takahashi}, {Wada}, \& {Yoshida}}]{Nomura+16}
{Nomura}, M., {Ohsuga}, K., {Takahashi}, H.~R., {Wada}, K., \& {Yoshida}, T. 2016, \pasj, 68, 16, \dodoi{10.1093/pasj/psv124}

\bibitem[{{Nord} {et~al.}(2004){Nord}, {Lazio}, {Kassim}, {Hyman}, {LaRosa}, {Brogan}, \& {Duric}}]{Nord+2004}
{Nord}, M.~E., {Lazio}, T. J.~W., {Kassim}, N.~E., {et~al.} 2004, \aj, 128, 1646, \dodoi{10.1086/424001}

\bibitem[{{Oh} {et~al.}(2018){Oh}, {Koss}, {Markwardt}, {Schawinski}, {Baumgartner}, {Barthelmy}, {Cenko}, {Gehrels}, {Mushotzky}, {Petulante}, {Ricci}, {Lien}, \& {Trakhtenbrot}}]{Oh+2018ApJS..235....4O}
{Oh}, K., {Koss}, M., {Markwardt}, C.~B., {et~al.} 2018, \apjs, 235, 4, \dodoi{10.3847/1538-4365/aaa7fd}

\bibitem[{Omeliukh {et~al.}(2024)Omeliukh, Barnier, \& Inoue}]{Omeliukh:2024fhg}
Omeliukh, A., Barnier, S., \& Inoue, Y. 2024.
\newblock \doarXiv{2411.09332}

\bibitem[{{Peretti} {et~al.}(2023){Peretti}, {Lamastra}, {Saturni}, {Ahlers}, {Blasi}, {Morlino}, \& {Cristofari}}]{Peretti+23}
{Peretti}, E., {Lamastra}, A., {Saturni}, F.~G., {et~al.} 2023, \mnras, 526, 181, \dodoi{10.1093/mnras/stad2740}

\bibitem[{Peretti {et~al.}(2023)Peretti, Peron, Tombesi, Lamastra, Saturni, Cerruti, \& Ahlers}]{Peretti:2023crf}
Peretti, E., Peron, G., Tombesi, F., {et~al.} 2023.
\newblock \doarXiv{2303.03298}

\bibitem[{{Ricci} {et~al.}(2017){Ricci}, {Trakhtenbrot}, {Koss}, {Ueda}, {Del Vecchio}, {Treister}, {Schawinski}, {Paltani}, {Oh}, {Lamperti}, {Berney}, {Gandhi}, {Ichikawa}, {Bauer}, {Ho}, {Asmus}, {Beckmann}, {Soldi}, {Balokovi{\'c}}, {Gehrels}, \& {Markwardt}}]{Ricci+2017ApJS..233...17R}
{Ricci}, C., {Trakhtenbrot}, B., {Koss}, M.~J., {et~al.} 2017, \apjs, 233, 17, \dodoi{10.3847/1538-4365/aa96ad}

\bibitem[{{Rybicki} \& {Lightman}(1986)}]{Rybicki+86}
{Rybicki}, G.~B., \& {Lightman}, A.~P. 1986, {Radiative Processes in Astrophysics}

\bibitem[{{Salvatore} {et~al.}(2024){Salvatore}, {Eichmann}, {Rodrigues}, {Dettmar}, \& {Becker Tjus}}]{Salvatore:2023zmf}
{Salvatore}, S., {Eichmann}, B., {Rodrigues}, X., {Dettmar}, R.~J., \& {Becker Tjus}, J. 2024, \aap, 687, A139, \dodoi{10.1051/0004-6361/202348447}

\bibitem[{{Sazonov} {et~al.}(2007){Sazonov}, {Revnivtsev}, {Krivonos}, {Churazov}, \& {Sunyaev}}]{Sazonov+2007}
{Sazonov}, S., {Revnivtsev}, M., {Krivonos}, R., {Churazov}, E., \& {Sunyaev}, R. 2007, \aap, 462, 57, \dodoi{10.1051/0004-6361:20066277}

\bibitem[{{Sazonov} {et~al.}(2004){Sazonov}, {Revnivtsev}, {Lutovinov}, {Sunyaev}, \& {Grebenev}}]{Sazonov+2004}
{Sazonov}, S.~Y., {Revnivtsev}, M.~G., {Lutovinov}, A.~A., {Sunyaev}, R.~A., \& {Grebenev}, S.~A. 2004, \aap, 421, L21, \dodoi{10.1051/0004-6361:20040179}

\bibitem[{{Schawinski} {et~al.}(2015){Schawinski}, {Koss}, {Berney}, \& {Sartori}}]{2015MNRAS.451.2517S}
{Schawinski}, K., {Koss}, M., {Berney}, S., \& {Sartori}, L.~F. 2015, \mnras, 451, 2517, \dodoi{10.1093/mnras/stv1136}

\bibitem[{{Shimizu} {et~al.}(2016){Shimizu}, {Mel{\'e}ndez}, {Mushotzky}, {Koss}, {Barger}, \& {Cowie}}]{Shimizu+16}
{Shimizu}, T.~T., {Mel{\'e}ndez}, M., {Mushotzky}, R.~F., {et~al.} 2016, \mnras, 456, 3335, \dodoi{10.1093/mnras/stv2828}

\bibitem[{{Skrutskie} {et~al.}(2006){Skrutskie}, {Cutri}, {Stiening}, {Weinberg}, {Schneider}, {Carpenter}, {Beichman}, {Capps}, {Chester}, {Elias}, {Huchra}, {Liebert}, {Lonsdale}, {Monet}, {Price}, {Seitzer}, {Jarrett}, {Kirkpatrick}, {Gizis}, {Howard}, {Evans}, {Fowler}, {Fullmer}, {Hurt}, {Light}, {Kopan}, {Marsh}, {McCallon}, {Tam}, {Van Dyk}, \& {Wheelock}}]{Skrutskie+2006}
{Skrutskie}, M.~F., {Cutri}, R.~M., {Stiening}, R., {et~al.} 2006, \aj, 131, 1163, \dodoi{10.1086/498708}

\bibitem[{{Tombesi} {et~al.}(2015){Tombesi}, {Mel{\'e}ndez}, {Veilleux}, {Reeves}, {Gonz{\'a}lez-Alfonso}, \& {Reynolds}}]{Tombesi+15}
{Tombesi}, F., {Mel{\'e}ndez}, M., {Veilleux}, S., {et~al.} 2015, \nat, 519, 436, \dodoi{10.1038/nature14261}

\bibitem[{{Tortosa} {et~al.}(2017){Tortosa}, {Marinucci}, {Matt}, {Bianchi}, {La Franca}, {Ballantyne}, {Boorman}, {Fabian}, {Farrah}, {Fuerst}, {Gandhi}, {Harrison}, {Koss}, {Ricci}, {Stern}, {Ursini}, \& {Walton}}]{Tortosa+17}
{Tortosa}, A., {Marinucci}, A., {Matt}, G., {et~al.} 2017, \mnras, 466, 4193, \dodoi{10.1093/mnras/stw3301}

\bibitem[{{Tsuboi} {et~al.}(2015){Tsuboi}, {Miyazaki}, \& {Uehara}}]{Tsuboi2015PASJ...67...90T}
{Tsuboi}, M., {Miyazaki}, A., \& {Uehara}, K. 2015, \pasj, 67, 90, \dodoi{10.1093/pasj/psv058}

\bibitem[{{Tueller} {et~al.}(2008){Tueller}, {Mushotzky}, {Barthelmy}, {Cannizzo}, {Gehrels}, {Markwardt}, {Skinner}, \& {Winter}}]{Tueller+2008}
{Tueller}, J., {Mushotzky}, R.~F., {Barthelmy}, S., {et~al.} 2008, \apj, 681, 113, \dodoi{10.1086/588458}

\bibitem[{{Tueller} {et~al.}(2010){Tueller}, {Baumgartner}, {Markwardt}, {Skinner}, {Mushotzky}, {Ajello}, {Barthelmy}, {Beardmore}, {Brandt}, {Burrows}, {Chincarini}, {Campana}, {Cummings}, {Cusumano}, {Evans}, {Fenimore}, {Gehrels}, {Godet}, {Grupe}, {Holland}, {Kennea}, {Krimm}, {Koss}, {Moretti}, {Mukai}, {Osborne}, {Okajima}, {Pagani}, {Page}, {Palmer}, {Parsons}, {Schneider}, {Sakamoto}, {Sambruna}, {Sato}, {Stamatikos}, {Stroh}, {Ukwata}, \& {Winter}}]{Tueller+2010}
{Tueller}, J., {Baumgartner}, W.~H., {Markwardt}, C.~B., {et~al.} 2010, \apjs, 186, 378, \dodoi{10.1088/0067-0049/186/2/378}

\bibitem[{{Wang} {et~al.}(2016){Wang}, {Liu}, {Qiu}, {Bai}, {Yang}, {Guo}, \& {Zhang}}]{Wang+2016}
{Wang}, S., {Liu}, J., {Qiu}, Y., {et~al.} 2016, \apjs, 224, 40, \dodoi{10.3847/0067-0049/224/2/40}

\bibitem[{{Wang} \& {Loeb}(2016{\natexlab{a}})}]{Wand&Loeb16a}
{Wang}, X., \& {Loeb}, A. 2016{\natexlab{a}}, Nature Physics, 12, 1116, \dodoi{10.1038/nphys3837}

\bibitem[{{Wang} \& {Loeb}(2016{\natexlab{b}})}]{Wang&Loeb16b}
---. 2016{\natexlab{b}}, \jcap, 2016, 012, \dodoi{10.1088/1475-7516/2016/12/012}

\bibitem[{{Weaver} {et~al.}(1977){Weaver}, {McCray}, {Castor}, {Shapiro}, \& {Moore}}]{1977ApJ...218..377W}
{Weaver}, R., {McCray}, R., {Castor}, J., {Shapiro}, P., \& {Moore}, R. 1977, \apj, 218, 377, \dodoi{10.1086/155692}

\bibitem[{{Yamada} {et~al.}(2024){Yamada}, {Sakai}, {Inoue}, \& {Michiyama}}]{Yamada+24}
{Yamada}, T., {Sakai}, N., {Inoue}, Y., \& {Michiyama}, T. 2024, \apj, 968, 116, \dodoi{10.3847/1538-4357/ad3a63}

\bibitem[{Yasuda {et~al.}(2024)Yasuda, Inoue, \& Kusenko}]{Yasuda:2024fvc}
Yasuda, K., Inoue, Y., \& Kusenko, A. 2024.
\newblock \doarXiv{2405.05247}

\bibitem[{{Ye} {et~al.}(2023){Ye}, {Hu}, {Tian}, {Chang}, {Chang}, {Cheng}, {Gao}, {Ge}, {Gong}, {Guo}, {Guo}, {He}, {Huang}, {Jiang}, {Jiang}, {Jing}, {Li}, {Li}, {Li}, {Li}, {Li}, {Liao}, {Lin}, {Lin}, {Liu}, {Liu}, {Liu}, {Miao}, {Mo}, {Morton-Blake}, {Peng}, {Sun}, {Tang}, {Tang}, {Tao}, {Tian}, {Wang}, {Wang}, {Wang}, {Wei}, {Wei}, {Wu}, {Xian}, {Xiang}, {Xu}, {Xue}, {Yang}, {Yang}, {Yu}, {Zeng}, {Zhang}, {Zhang}, {Zhang}, {Zhang}, {Zhi}, {Zhong}, {Zhou}, {Zhu}, \& {Zhuang}}]{Ye+2023}
{Ye}, Z.~P., {Hu}, F., {Tian}, W., {et~al.} 2023, Nature Astronomy, 7, 1497, \dodoi{10.1038/s41550-023-02087-6}

\bibitem[{{Zoglauer} {et~al.}(2021){Zoglauer}, {Siegert}, {Lowell}, {Mochizuki}, {Kierans}, {Sleator}, {Hartmann}, {Lazar}, {Gulick}, {Beechert}, {Roberts}, {Tomsick}, {Leising}, {Pellegrini}, {Boggs}, \& {Brandt}}]{Zoglauer+2021}
{Zoglauer}, A., {Siegert}, T., {Lowell}, A., {et~al.} 2021, arXiv e-prints, arXiv:2102.13158, \dodoi{10.48550/arXiv.2102.13158}

\end{thebibliography}
\bibliographystyle{aasjournal}



\end{sloppypar}
\end{document}